\documentclass{ws-ijmpa}
\usepackage[super,compress]{cite}

\newcommand{\re}{\text{Re }}
\newcommand{\im}{\text{Im }}

\newcommand{\bra}[1]{\langle  #1  |}
\newcommand{\ket}[1]{|  #1  \rangle}
\newcommand{\bbra}[1]{\langle  #1  }
\newcommand{\kket}[1]{  #1  \rangle}

\newcommand{\PDG}{Beringer:1900zz}

\begin{document}

\markboth{Tetsuo Hyodo}
{Structure and Compositeness of Hadron Resonances}

%
\catchline{}{}{}{}{}
%

\title{STRUCTURE AND COMPOSITENESS OF \\
HADRON RESONANCES
}

\author{TETSUO HYODO}

\address{Yukawa Institute for Theoretical Physics, \\
Kyoto University, Kyoto 606-8502, Japan\\
hyodo@yukawa.kyoto-u.ac.jp}

\maketitle

\begin{history}
\received{30 September 2013}
\revised{7 October 2013}
\end{history}

\begin{abstract}
The structure of the hadron resonances attracts much attention, in association with the recent observations of various exotic hadrons which do not fit well in the conventional picture. These findings urge us to consider various new configurations such as the multiquark states and the hadronic molecules. However, it is a subtle problem to define a proper classification scheme for the hadron structure, and the nonzero decay width of the hadron resonances makes the analysis complicated. In this paper, we summarize the recent developments in the studies of the structure of the hadron resonances, focusing on the notion of the compositeness in terms of the hadronic degrees of freedom.

\keywords{Exotic hadrons; hadron structure; compositeness; elementariness; resonances.}
\end{abstract}

\ccode{PACS numbers: 03.65.Nk, 14.20.Gk, 21.45.-v}

\section{Introduction}	

Strong interaction is governed by Quantum Chromodynamics (QCD) of quarks and gluons. Because of the color confinement, the low-energy spectrum of QCD consists of plenty of color-singlet hadrons.\cite{\PDG} In the traditional picture of naive constituent quark models, the mesons are described as the quark--antiquark pair $(q\bar{q})$ and its internal excitations, and the baryons as the three-quark configuration $(qqq)$ with its excitations. Recent developments in the high-energy experimental facilities (Belle, BaBar, CLEO, BESIII, LHCb, etc.) identify a bunch of new hadrons, especially in the heavy (charm and bottom) quark sector.\cite{Brambilla:2010cs} At the same time, the accumulation of the precise data in the low-energy exclusive measurements (LEPS, CLAS, HADES, COSY, etc.) reveals the new aspects of the light (up, down and strange) hadrons, such as the $\Lambda(1405)$ resonance.\cite{Hyodo:2011ur} These activities clearly establish the existence of the hadrons which do not fit well in the predictions of the conventional picture. These are called exotic hadrons.\footnote{In literature, the word ``exotic hadrons'' is sometimes reserved for the manifestly exotic states which require more than three valence quarks. Here, we simply use it for the hadrons with non-conventional properties.}

In order to explain the extraordinary nature of the exotic hadrons, many kinds of new configurations have been proposed for the internal structure of the exotic hadrons. Typical configurations can be classified into three categories as follows.
\begin{itemlist}
\item \textit{Multi-quark hadrons}: Hadrons with four or more valence quarks, e.g. the mesons made of four quarks (tetraquarks, $qq\bar{q}\bar{q})$ and the baryons made of five quarks (pentaquarks, $qqqq\bar{q})$.\cite{Jaffe:1976ig,Jaffe:1976ih,Jaffe:2004ph} 
\item \textit{Hadronic molecules}: Loosely-bound two or more hadron systems, e.g. the mesons with the meson--meson molecular structure and the baryons with the meson--baryon molecular structure.\cite{Tornqvist:1991ks} 
\item \textit{Gluonic hybrid}: The hadrons with the valence gluon component, e.g. $q\bar{q}g$, $qqqg$,\,\dots, or 
the mesons entirely made by gluons (glueball, $gg$).\cite{Klempt:2007cp}
\end{itemlist}

As far as the color confinement is concerned, all these states are allowed to exist (with appropriate color configurations to form the singlet in total). Moreover, because the exotic hadrons are usually found in the excited spectrum, it is natural to expect the creation of $q\bar{q}$ pairs, gluons and virtual mesons as an excitation mode on top of the ground states. Thus, the exotic hadrons are expected to manifest the new forms of the hadrons, and hence the study of them helps us to learn how the hadrons are constructed from the highly nontrivial dynamics of the strong interaction. In this way, the study of the hadron structure deserves a good testing ground for the low-energy nonperturbative dynamics of QCD.

On the other hand, as we try to clarify the hadron structure, we will encounter some subtle issues in the discussion. For instance, how can we distinguish the four-quark state from the meson--meson molecule? What is the wave function of an exotic hadron with a finite decay width? In literature, these difficulties have been realized (although sometimes they are disregarded for simplicity), but there is no broad consensus for the proper measure of the hadron structure. It is therefore desirable to have a general classification scheme, which does not rely upon the specific models. This is the issue we would like to address in this paper.

In the following, we first summarize the subtleties of the definition of the  hadron structure in some detail, and try to establish an appropriate strategy for this problem in Sec.~\ref{sec:strategy}. Based on this strategy, we discuss the structure of stable bound states from the viewpoint of the compositeness of the hadrons in sections~\ref{sec:bound} and \ref{sec:resonance}. For a stable bound state, the compositeness is well-defined and normalized. Moreover, for a weakly bound state, the compositeness can be related to the experimental observables in a model-independent manner. However, it is not straightforward to generalize this approach to resonances. We show recent efforts of the application of the compositeness to resonances in Sec.~\ref{sec:resonance}. In Sec.~\ref{sec:other}, we briefly introduce other approaches to the hadron structure, which are complementary to the compositeness approach. The last section is devoted to summary.

\section{Strategy to Study the Structure of Hadrons}\label{sec:strategy}

We discuss the subtleties and difficulties in the discussion of the structure of hadrons. Some of them are well known, but others are not thoroughly considered in literature. Examining these difficulties, we would like to establish the conditions for the desirable formulation of the hadron structure.

\subsection{Classification scheme of the hadron structure}

We have introduced several different configurations for the description of the exotic hadrons. Naively, one may ask whether a given exotic hadron is the $q\bar{q}$ state \textit{or} the tetraquark state \textit{or} the mesonic molecule, and so on. This is not a proper question, because all possible configurations with the same quantum numbers should mix with each other in the quantum field theory. A hadronic state should be written as a superposition of all possible configurations and our task is to pin down the \textit{dominant} configuration over the other components. 

Let us take an example of the $\Lambda(1405)$ baryon, a negative parity excited state with the strangeness $S=-1$ and the isospin $I=0$.\cite{Hyodo:2011ur} In the conventional constituent quark model, this state is an orbital excitation of one of the three quarks of the ground state $\Lambda$ made of $uds$.\cite{Isgur:1978xj} On the other hand, $\Lambda(1405)$ can be well described in the coupled-channel meson--baryon scattering model, which indicates the quasi-bound $\bar{K}N$ molecule picture for the $\Lambda(1405)$ resonance.\cite{Dalitz:1967fp} Moreover, it is also possible to consider a five-quark configuration ($uds$ plus $q\bar{q}$) where no orbital excitation is needed due to the negative parity of the antiquark.\cite{Strottman:1979qu} In this way, the structure of $\Lambda(1405)$ may be schematically decomposed as
\begin{align}
    \ket{\Lambda(1405)}
    = 
    N_{3q}\ket{uds}
    +
    N_{5q}\ket{uds\ q\bar{q}}
    +
    N_{\bar{K}N}\ket{\bar{K}N}
    +\cdots 
     .
    \label{eq:L1405}
\end{align}
The dominant configuration is the one with the largest $N_{i}$ among others. At first glance, this strategy sounds reasonable. 

However, we should remember that a meaningful decomposition requires the orthogonality of the expansion basis. Because the ground state mesons and baryons should be dominated by the $q\bar{q}$ and $qqq$ configurations, it is plausible that the five-quark component has an overlap with the meson--baryon state, namely,
\begin{align}
    \bra{uds\ q\bar{q}}\kket{\bar{K}N}
    \neq 0\; .
\end{align}
This means that the basis is not orthogonal and the decomposition~\eqref{eq:L1405} is not adequate. In fact, two kinds of bases are mixed in Eq.~\eqref{eq:L1405}; the states with the quarks and the states with the hadrons. To begin with, we should choose the relevant degrees of freedom, otherwise the orthogonality of the basis is not clearly ensured.

\subsection{Number of quarks or number of hadrons?}

Let us first examine the basis in terms of quarks (and gluons, if necessary). In this case, the expansion of the $\Lambda(1405)$ may be 
\begin{align}
    \ket{\Lambda(1405)}
    =N_{3q}\ket{uds}
    +
    N_{5q}\ket{uds\ q\bar{q}}
    +
    N_{7q}\ket{uds\ q\bar{q}\ q\bar{q}}
    +\cdots .
    \label{eq:decompquark}
\end{align}
In QCD, the number of quarks $n_{q}$ is conserved as a consequence of the vectorial U(1) symmetry. If we allow the existence of the antiquarks, what is conserved is the net quark number of the system, $n_{q}-n_{\bar{q}}$ where $n_{\bar{q}}$ is the number of the antiquarks. We should however notice that the \textit{sum of the number of quarks and the number of antiquarks $(n_{q}+n_{\bar{q}})$ is not a conserved quantum number}. In other words, the QCD Hamiltonian can create or annihilate any number of $q\bar{q}$ pairs without modifying the quantum number of the system. This means that the three-quark state is not orthogonal to the five-quark state,
\begin{align}
    \bra{qqq}\kket{qqq\ q\bar{q}}
    \neq 0\; .
\end{align}
In this way, counting the number of quarks and antiquarks in a hadron is not a good classification scheme\footnote{An exception of this argument is the infinite momentum frame where the $q\bar{q}$ pairs are separated from the valence component and hence the Fock space expansion is well defined.\cite{Diakonov:2004as} }. This is because the quarks and gluons are not the asymptotic states of the QCD vacuum. 

Instead, it is possible to work with the hadronic degrees of freedom. In this case, the expansion may be
\begin{align}
    \ket{\Lambda(1405)}
    =N_{B}\ket{B}
    +
    N_{BM}\ket{BM}
    +
    N_{BMM}\ket{BMM}
    +\cdots .
    \label{eq:decomphadron}
\end{align}
An important difference from Eq.~\eqref{eq:decompquark} is that each component is now written in terms of the hadronic degrees of freedom which are the asymptotic states of the QCD vacuum. Let us demonstrate the virtue of this fact, following the discussion in Ref.~\citen{Hanhart:2007cm}. Consider the two-point correlation function of an operator $\mathcal{O}$ which can create $\Lambda(1405)$. The intermediate states can be expanded by the basis of Eq.~\eqref{eq:decomphadron} as
\begin{align}
    \bra{0}\mathcal{O}(x)\mathcal{O}^{\dag}(y)\ket{0}
    &= 
    \sum_{n}\bra{0}\mathcal{O}(x)\ket{n}\bra{n}\mathcal{O}^{\dag}(y)\ket{0}\nonumber \\
    &= 
    \bra{0}\mathcal{O}(x)\ket{B}\bra{B}\mathcal{O}^{\dag}(y)\ket{0} \nonumber \\
    &\quad +\bra{0}\mathcal{O}(x)\ket{BM}\bra{BM}\mathcal{O}^{\dag}(y)\ket{0}
    +\cdots .
    \label{eq:correlation}
\end{align}
The first term corresponds to the propagation of a single hadron as a whole, the second term represents the contribution from a two-hadron state $\ket{BM}$, and so on. The Fourier transform of the first term of the correlation function~\eqref{eq:correlation} is an analytic function of the energy variable. On the other hand, the contribution from the second term shows a non-analytic behavior at the threshold of the $BM$ state $\propto \sqrt{E}$. Because the correlation function has an imaginary part above the threshold, the branch point exists at the threshold. Thus, it may be possible to distinguish each component, by using the different analytic structure. Note that this distinction is only possible with the asymptotic degrees freedom, otherwise the two-particle state does not exhibit the nonanalyticity. Now, we have reached the first conclusion; \textit{the structure of hadrons should be discussed in terms of the hadronic (in general, asymptotic) degrees of freedom}.

\subsection{Model space and the CDD pole contributions}

Let us focus on a different aspect. A conventional strategy for the structure of hadrons is to compare the model prediction with experimental data, such as the mass spectra and the decay properties. For instance, the $\Lambda(1405)$ resonance is considered to be an exotic hadron rather than the simple $qqq$ state, because the light mass of $\Lambda(1405)$ is not well reproduced in the conventional three-quark model. This is fine as a first step, but we notice that this criterion loses power when the models are very much sophisticated. Suppose that we try to improve the quark model by introducing various interaction potentials with many adjustable parameters. At the end of the day, we may be able to describe the $\Lambda(1405)$ resonance without disturbing the predictions of other hadrons, but with hundreds of parameters. Clearly, the success of such a fine-tuned model does not convince us to regard the $\Lambda(1405)$ resonance as the three-quark state. It is more likely that the other components (such as the meson--baryon molecules) are effectively renormalized into the parameters of the complicated potential. In this sense, a good model description of a hadron does not necessarily mean that the structure of the hadron is dominated by the component of the model space. The same discussion is equally applied to the dynamical scattering model. We have mentioned that the $\Lambda(1405)$ resonance can be dominated by the meson--baryon molecule component, because of the success of the meson--baryon scattering model. Again, this is fine as a first orientation, but is not always relevant for the same reasons. In some cases, the calculation can be systematically improved with an well-founded expansion scheme, such as chiral perturbation theory.\cite{Gasser:1983yg,Gasser:1985gg} Unfortunately, even in this case, the contribution other than the model space can be hidden in the model parameters. For instance, the effect of the vector mesons is included in the low-energy constants of the higher-order terms in chiral perturbation theory whose explicit degrees of freedom are the pseudoscalar mesons~\cite{Ecker:1988te}. This again indicates the vagueness of the connection of the model space and the structure of the described hadrons.

Related to this issue, it is important to recall the old discussion of the Castillejo--Dalitz--Dyson (CDD) ambiguity.\cite{Castillejo:1956ed} Originally, it was shown that there is an ambiguity of adding any number of poles in the denominator function in the $N/D$ method\cite{Chew:1960iv,Bjorken:1960zz} (see also a recent discussion on this ambiguity in connection with the crossing symmetry\cite{Mcleod:2013qua}). The CDD pole contribution was later interpreted as the effect from an unstable elementary particle, because they are equivalent to each other within the $N/D$ framework.\cite{PR124.264} This means that the poles of the scattering amplitude can be generated not only from the dynamical effect by the interaction potential, but also from the subtraction of the dispersion integral. The former states are interpreted as dynamically generated states within the model space, while the latter states are introduced from the outside of the model space. In fact, it is explicitly demonstrated for the energy-dependent interaction that the subtraction constant in the loop function can play a role of the pole term in the interaction kernel.\cite{Hyodo:2008xr} In this way, the prepared model space does not always correspond to the structure of the generated states in the scattering theory.

In summary, we argue that the origin of the hadrons is not exclusively attributed to the degrees of freedom of the models, and the other contributions can be hidden in the model parameters. In this sense, all these subtleties can be traced back to the lack of \textit{the model-independent measure of the hadron structure}. For instance, if the hadron structure is related to the experimental observables, it is a model-independent discussion. Once we establish a model-independent quantity which characterizes the hadron structure, it is meaningful to calculate that quantity in various models, because the result can be compared with other models. In principle, such quantity can as well be calculated in the lattice QCD simulation. The model-independent nature is the second feature to be equipped in the desired framework of the hadron structure.

\subsection{Difficulty of resonances}

Yet another difficulty arises from the strong decay of the hadron excited states. The candidates for the exotic hadrons are identified in the spectrum of the excited states, most of them have a nonzero decay width via the strong interaction. Although we have not explicitly considered the effect of the width so far, it induces further subtleties of the discussion of the hadron structure. For instance, we write the decomposition of the $\Lambda(1405)$ resonance in Eqs.~\eqref{eq:L1405}, \eqref{eq:decompquark} and \eqref{eq:decomphadron}, but the ``state vector'' in the left-hand side should be carefully constructed.

In the scattering theory, the resonance states are identified by the poles of the scattering amplitude in the second (unphysical) Riemann sheet of the complex energy plane.~\cite{Taylor} This is a natural generalization of the stable bound states which are expressed by the poles on the real axis of the first (physical) Riemann sheet. Through the comparison with the Breit--Wigner parametrization of the resonance amplitude, the real (imaginary) part of the pole position is interpreted as the mass (half width) of the resonance state. 

From this viewpoint, the resonance states can be regarded as the eigenstates of the Hamiltonian with a complex eigenvalue. It is however not straightforward to construct the state vector with complex eigenvalues, because the eigenvalues of an Hermite operator must be real in the normal Hilbert space spanned by the square integrable functions. Mathematically, complex eigenvalues can be achieved by the Gamow vectors in the rigged Hilbert space.\cite{Gamow:1928zz,PTP33.1116,Berggren:1968zz,Bohm:1981pv,Kukulin} In this case, however, the expectation value of an Hermite operator becomes complex for the Gamow states.\cite{PL33B.547,PLB373.1} In fact, the mean squared radii and the form factors of $\Lambda(1405)$ are obtained as complex numbers, when $\Lambda(1405)$ is treated as the resonance in the scattering amplitude.\cite{Sekihara:2008qk,Sekihara:2010uz} The interpretation of the complex-valued quantities is not straightforward.

In any event, it is inevitable to consider the resonances for the discussion of the structure of the exotic hadrons. The ideal classification scheme should be applicable to the resonance states, with a natural interpretation of the obtained results.

\subsection{Conditions for a proper classification scheme}\label{subsec:conditions}

We have discussed the difficulties in the study of the structure of the exotic hadrons. In view of these discussions, we can summarize the desired feature of the proper classification scheme for the hadron structure as follows.
\begin{itemize}

\item The structure of the hadrons should be classified in terms of the \textit{hadronic degrees of freedom}, and 

\item the scheme should be \textit{model independent} and hopefully related to the \textit{experimental observables}, and 

\item the \textit{unstable resonance states} can be treated in the same way with the stable bound states with providing a meaningful interpretation of the results.

\end{itemize}
As a matter of fact, there is no framework which completely satisfies all these requirements, especially for the last point. In Sec.~\ref{sec:bound}, we introduce one of the best approaches for the hadron structure, utilizing the compositeness of hadrons. This method defines the compositeness in terms of the hadronic degrees of freedom, which can be model-independently related to the experimental observables in the weak-binding limit. Its generalization to the resonances is discussed in Sec.~\ref{sec:resonance}.

\section{Compositeness of Stable Bound States}\label{sec:bound}

In the early 1960s, one of the central problems in the particle physics was to distinguish between elementary and composite particles. It is eventually realized that the field renormalization constant $Z$ is useful for this distinction~\cite{Salam:1962ap,PTP29.877,Weinberg:1962hj,Weinberg:1963zz,Weinberg:1964zz,PR136.B816}. The constant $Z$ represents the probability of finding the elementary component in the physical state, and the quantity $1-Z$ stands for the compositeness of the state. In particular, it was shown that the field renormalization constant of a weakly bound state can be related to the threshold parameters (the scattering length and the effective range) in a model-independent way~\cite{Weinberg:1965zz}. With this method, the deuteron was shown to be dominated by the two-nucleon composite structure. This conclusion is of course naturally expected from the study of the nuclear force. But the conclusion of Ref.~\citen{Weinberg:1965zz} is drawn without any information on the $NN$ potential and the wave function of the deuteron. Strictly speaking, the potentials and wave functions are not observables and can be simultaneously changed by the unitary transformation with keeping the physical observables unchanged. In the same way, the field renormalization constant in general is also a scheme-dependent quantity. The virtue of the analysis in Ref.~\citen{Weinberg:1965zz} is that the field renormalization constant is model-independently related to the experimental observables, by taking the weak-binding limit. This is suitable for our strategy described in the previous section.

In the following, we review the formulation of the compositeness using the field renormalization constant in the case of the stable bound states, basically following Ref.~\citen{Weinberg:1965zz}. For simplicity, we consider the single-channel scattering with one bound state, and suppress the indices for possible internal degrees of freedom (such as spin, isospin, etc.) which can be amended straightforwardly. We first describe the construction of the basis in Subsecs.~\ref{subsec:standard} and \ref{subsec:bare}, and define the field renormalization constant and the compositeness in Subsec.~\ref{subsec:Zdef}. A concrete examp example of this formulation in quantum field theory is given in Subsec.~\ref{subsec:example} for illustration. The expression of the compositeness by the scattering amplitude is then given in Subsec.~\ref{subsec:amp}. Up to this point, all the results are exact. Next we consider the weak-binding limit in Subsec.~\ref{subsec:weakbinding}, paying attention to the neglected terms in the weak-binding expansion. The model-independent relation to the threshold parameters are obtained here. In Subsec.~\ref{subsec:interpretation}, we provide a discussion for the interpretation of the result. The case of the physical deuteron is analyzed in Subsec.~\ref{subsec:deuteron}. In Subsec.~\ref{subsec:multi}, we present the generalization to the coupled-channel scattering. 

\subsection{Standard basis for the scattering problem}
\label{subsec:standard}

We consider a nonrelativistic quantum system described by a Hamiltonian $H$ with $\hbar=1$. We decompose the Hamiltonian into the kinetic term $K$ and the potential term $V$ as 
\begin{align}
    H &= K + V\; ,
    \label{eq:decompose1} \\
    K&=\frac{\hat{\bm{p}}^{2}}{2\mu}\; ,
\end{align}
where $\hat{\bm{p}}$ is the momentum operator and $\mu$ is the reduced mass of the system. We denote the eigenstates of the kinetic term operator $K$ as $\ket{\bm{p}}$, which has the following property
\begin{align}
    K \ket{\bm{p}}
    &=E_{p}\ket{\bm{p}}\; , \\
    E_{p}
    &\equiv \frac{\bm{p}^{2}}{2\mu}\; ,
\end{align}
where $p=|\bm{p}|$. In the coordinate space, $\ket{\bm{p}}$ represents the plane wave, $\bra{\bm{x}}\kket{\bm{p}}=e^{i\bm{p}\cdot\bm{x}}/(2\pi)^{3/2}$. The normalization of the eigenvectors is given by
\begin{align}
    \bbra{\bm{p}^{\prime}}\ket{\bm{p}}
    &=\delta(\bm{p}^{\prime}-\bm{p})\; ,
\end{align}
and the completeness relation ensures that
\begin{align}
    1
    &=
    \int d^{3}\bm{p} \ket{\bm{p}}\bra{\bm{p}}\; .
\end{align}
Because the norm is infinite, the state vector $\ket{\bm{p}}$ is formally referred to as an ``improper'' vector.

Now, we consider the eigenstates of the full Hamiltonian. The asymptotic (in and out) scattering states are defined as the images of $\ket{\bm{p}}$ under the M\o ller operators $\Omega_{\pm}=\lim_{t\to\mp\infty}e^{iHt}e^{-iH^{0}t}$:
\begin{align}
    \ket{\bm{p},\pm} \equiv \Omega_{\pm}\ket{\bm{p}}\; .
\end{align}
With the help of the intertwining relation, we can show that
\begin{align}
    H\ket{\bm{p},\pm} 
    &=E_{p}\ket{\bm{p},\pm}\; , \\
    \bbra{\bm{p}^{\prime},\pm}\ket{\bm{p},\pm}
    &=\delta(\bm{p}^{\prime}-\bm{p})\; .
\end{align}
Because the M\o ller operators are isometric but not necessarily be unitary, it is possible to include a discrete eigenstate $\ket{B}$ with an energy $E=-B, B>0$ in addition to the scattering states:
\begin{align}
    H\ket{B} 
    &=-B\ket{B}\; .
\end{align}
Namely, $\ket{B}$ represents a two-body bound state with the binding energy $B$. Because the bound state wave function is square integrable, $\ket{B}$ can be normalized as 
\begin{align}
    \bra{B}\kket{B}
    &= 1\; .
\end{align}
It is shown for reasonably ``well-behaved'' potentials\footnote{For instance, conditions for a spherical potential $V(r)$ are $V(r)=\mathcal{O}(r^{-3-\epsilon})$ as $r\to \infty$, $V(r)=\mathcal{O}(r^{-2+\epsilon})$ as $r\to 0$, and continuous for $0<r<\infty$, except at a finite number of finite discontinuities.} that the subspace spanned by the asymptotic scattering states are orthogonal to that spanned by the bound states and the model space is asymptotically complete.~\cite{Taylor} We then have the relations
\begin{align}
    \bra{B}\kket{\bm{p},\pm}
    &=0 \; , \\
    1
    &=\int d\bm{p}\ket{\bm{p},\pm }\bra{\bm{p},\pm }
    +\ket{B}\bra{B}\; .
\end{align}
Namely, the scattering states $\ket{\bm{p},\pm}$ and the bound state $\ket{B}$ form an orthogonal and complete basis for the Hamiltonian $H$. 

\subsection{Introduction of the bare state}\label{subsec:bare}

For the discussion of the compositeness, we define the ``bare Hamiltonian'' $H_{0}$ which has a discrete level of the ``bare bound state'' $\ket{B_{0}}$ as an eigenstate. Namely, we decompose the Hamiltonian as
\begin{align}
    H
    = H_{0}+V 
    \label{eq:decomposition}
\end{align}
and consider that the orthogonal basis for the bare Hamiltonian $H_{0}$ is
\begin{align}
    H_{0}\ket{\bm{p}}
    &= E_{p}\ket{\bm{p}}\; ,\quad
    H_{0}\ket{B_{0}}
    = -B_{0}\ket{B_{0}}\; , 
    \label{eq:eigen0}\\
    \bra{\bm{p}}\kket{\bm{p}^{\prime}}
    &=\delta(\bm{p}^{\prime}-\bm{p})\; ,
    \quad 
    \bra{B_0}\kket{B_0}=1 \; ,\quad
    \bra{B_0}\kket{\bm{p}}
    =0 
    \label{eq:orthogonality0}\; ,
    \\
    1 
    &= \ket{B_0}\bra{B_0} + \int d\bm{p} \ket{\bm{p}}\bra{\bm{p}}
    \; ,
    \label{eq:completeness0}
\end{align}
where $B_{0}>0$. One may wonder about the origin of the bare bound state $\ket{B_{0}}$. In general, the origin of $\ket{B_{0}}$ stems from the dynamics at a deeper level. In the following, we present a schematic illustration of the origin of $\ket{B_{0}}$ in two different ways. An explicit model Hamiltonian is also constructed in Subsec.~\ref{subsec:example}. 

One possible interpretation is to regard $\ket{B_{0}}$ as a bound state of the higher energy channel. Consider a two-channel problem,
\begin{align}
    H
    &= K_{1}+V_{1}+K_{2}+V_{2}+V_{\rm mix}
    \; ,
    \label{eq:2channel}
\end{align}
where $K_{i}$ and $V_{i}$ are the kinetic and potential terms in the channel $i=1,2$ and $V_{\rm mix}$ represents the mixing potential among the two channels. Performing the Feshbach projection method\cite{Feshbach:1958nx,Feshbach:1962ut}, we construct an equivalent Hamiltonian acting only on the channel $1$ as
\begin{align}
    H
    &= K_{1}+V_{1}+V_{\rm eff}
    \; ,
\end{align}
where the effective potential is schematically given by $V_{\rm eff}=V_{\rm mix}(E-K_{2}-V_{2})^{-1}V_{\rm mix}$. If the interaction $V_{2}$ supports a bound state $\ket{B_{0}}$, we can define
\begin{align}
    H_{0}
    &= K_{1}+V_{\rm eff}P_{B_{0}}
    \; , \\
    V
    &=V_{1}+V_{\rm eff}(1-P_{B_{0}})\; ,
\end{align}
where $P_{B_{0}}$ is the projection operator to the $\ket{B_{0}}$ state. In this way, the eigenstates of $H_{0}$ are the scattering states from $K_{1}$ and the bound state from $V_{\rm eff}P_{B_{0}}$. 

We can also consider the system where the asymptotic degrees freedom are different from those in the original Hamiltonian. This is indeed the case in QCD; because of the color confinement, the asymptotic degrees of freedom in the low-energy vacuum are the hadrons, not the quarks and gluons in the QCD Hamiltonian. In such cases, the role of the ``potential'' term $V$ is twofold;
\begin{itemize}

\item $V$ forms the color-singlet hadrons from the quarks and gluons, and  

\item $V$ produces the inter-hadron forces.

\end{itemize}
Suppose that we decompose the QCD potential $V$ into $V_{\rm conf}$ which is responsible for the confinement and $V_{\rm int}$ for the inter-hadron interactions. The full Hamiltonian is then given by
\begin{align}
    H = K_{\rm QCD} + V_{\rm conf} + V_{\rm int}\; .
    \label{eq:decompose2}
\end{align}
Here $K_{\rm QCD}$ represents the kinetic terms of quarks and gluons. We define the bare Hamiltonian as the system only with $V_{\rm conf}$, 
\begin{align}
    H_{0} &= K_{\rm QCD} + V_{\rm conf}\; , \\
    V &= V_{\rm int}\; .
\end{align}
Now the asymptotic eigenstates of $H_{0}$ are the hadrons, but their interactions are switched off $(V_{\rm int}=0)$. In other words, $H_{0}$ describes the system of free non-interacting hadrons. When $V_{\rm conf}$ produces the hadrons in the scattering channel and the bare bound state, the eigenstates of $H_{0}$ are given by $\ket{\bm{p}}$ and $\ket{B_{0}}$.

We note the essential ambiguity of the choice of the bare Hamiltonian. In Eq.~\eqref{eq:2channel}, the dynamics in channel 2 can be arbitrarily chosen. In Eq.~\eqref{eq:decompose2}, the decomposition of $V_{\rm conf}$ and $V_{\rm int}$ is not trivial. In addition, we may as well consider the case with multiple bare bound states. At this point, we mention that it is our choice to use the basis~\eqref{eq:completeness0} for the investigation of the structure of the physical bound state. 

\subsection{Field renormalization constant and the compositeness}\label{subsec:Zdef}

\begin{figure}[tpb]
\centerline{
\psfig{file=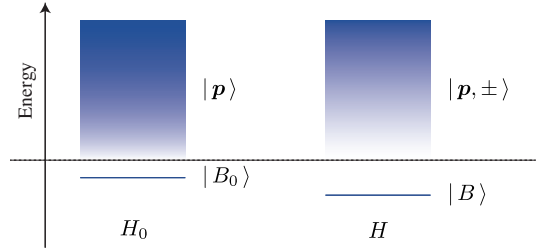,width=7cm}
}
\vspace*{8pt}
\caption{Schematic illustration of the spectra of the free Hamiltonian $H_{0}$ (left) and the full Hamiltonian $H$ (right) in the single channel scattering. \label{fig:spectrum1}}
\end{figure}

Introducing the bare state, we are now in a position to define the field renormalization constant and the compositeness. We summarize the eigenstates of the full Hamiltonian $H$ as
\begin{align}
    H\ket{\bm{p},\pm}
    &= E_{p}\ket{\bm{p},\pm}\; ,\quad
    H\ket{B}
    = -B\ket{B}\; , 
    \label{eq:eigenfull} \\
    \bra{\bm{p},\pm}\kket{\bm{p}^{\prime},\pm}
    &=\delta(\bm{p}^{\prime}-\bm{p})\; ,
    \quad 
    \bra{B}\kket{B}=1\;  ,\quad
    \bra{B}\kket{\bm{p},\pm}
    =0 \; ,
    \\
    1 
    &= \ket{B}\bra{B} + \int d\bm{p} \ket{\bm{p},\pm}\bra{\bm{p},\pm}
    \; .
    \label{eq:completenessfull}
\end{align}
This basis and that in Eqs.~\eqref{eq:eigen0}, \eqref{eq:orthogonality0} and \eqref{eq:completeness0} establish the foundation of the discussion. We use the basis of the bare Hamiltonian to decompose the wave function of the physical bound state $\ket{B}$. The eigenstates of $H_{0}$ and $H$ are schematically depicted in Fig.~\ref{fig:spectrum1}.

The field renormalization constant $Z$ is introduced as the overlap of the physical bound state $\ket{B}$ and the bare bound state $\ket{B_{0}}$:
\begin{align}
    Z
    & \equiv |\bra{B_0}B\, \rangle |^2\;  ,
    \label{eq:Zdef}
\end{align}
which represents the probability of finding the bound state $B$ in the bare state $B_0$. In other words, $Z$ represents the elementariness of the physical bound state $\ket{B}$.\footnote{The word ``elementary'' is used to mean the contribution from $\ket{B_{0}}$. Once we start discussion by specifying the expansion basis~\eqref{eq:completeness0}, there is no way to ask the origin of $\ket{B_{0}}$.} In the same way, we define the compositeness $X$ as the overlap of the bound state $\ket{B}$ and the scattering states\footnote{In literature, the compositeness is written as $1-Z$ which is equivalent to $X$ because of the normalization~\eqref{eq:normalization}.}
\begin{align}
    X
    & \equiv \int d\bm{p} |\bra{\bm{p}}B\, \rangle |^2\;  .
    \label{eq:Xdef}
\end{align}
Using the completeness relation~\eqref{eq:completeness0} and the normalization $\bra{B}\kket{B}=1$, we obtain the normalization of $Z$ and $X$ as
\begin{align}
    1 
    & =Z+X\;  .\label{eq:normalization} 
\end{align}
This means that the elementariness $Z$ and the compositeness $X$ are exclusive with each other. Together with the non-negativeness of Eqs.~\eqref{eq:Zdef} and \eqref{eq:Xdef}, Eq.~\eqref{eq:normalization} leads to
\begin{align}
    0
    & \leq Z\leq 1\;  , \\
    0
    & \leq X\leq 1\;  . 
\end{align}
This shows that both $Z$ and $X$ are the normalized quantities. This fact ensures that the probabilistic interpretation is possible; $Z$ ($X$) represents the fraction of the physical bound state $\ket{B}$ as the elementary (composite) component. 

It follows from the eigenvalue equations~\eqref{eq:eigen0} and \eqref{eq:eigenfull} that
\begin{align}
    \bra{\bm{p}}H_{0}+V\ket{B}
    &=\bra{\bm{p}}E_{p}+V\ket{B}
    = 
    -B\bra{\bm{p}}\kket{B}\;  ,
\end{align}
so the compositeness can be written as
\begin{align}
    X
    &= 
    \int d\bm{p}
    \frac{|\bra{\bm{p}}V\ket{B}|^2}{(E_{p}+B)^2}\;  ,
    \label{eq:Xgeneral}
\end{align}
where the matrix element $\bra{\bm{p}}V\ket{B}$ represents the transition form factor of the bound state $\ket{B}$ to the scattering state with a momentum $\bm{p}$ through the interaction $V$. From now on, we focus on the $s$-wave bound states. In this case, the form factor $\bra{\bm{p}}V\ket{B}$ does not depend on the angular variable and can be specified by the magnitude of the momentum $p$, or equivalently, by the energy variable $E_{p}=\bm{p}^{2}/2\mu$. We thus define the transition form factor for the $s$-wave bound state
\begin{align}
    \bra{\bm{p}}V\ket{B}\equiv G(E_{p}) 
    \label{eq:FF}
\end{align}
and rewrite Eq.~\eqref{eq:Xgeneral} as
\begin{align}
    X
    &=1-Z=4\pi\sqrt{2\mu^3}
    \int_{0}^{\infty} dE\frac{\sqrt{E}
    |G(E)|^2}{(E+B)^2} . 
    \label{eq:Xexact}
\end{align}
This is the exact expression of the compositeness $X$ for an $s$-wave bound state. If we know the binding energy $B$ and the form factor $G(E)$, we can calculate the compositeness $X$.

\subsection{Field theoretical example}\label{subsec:example}

The above discussion is given by the general expressions with the abstract state vectors. It is instructive to demonstrate the same discussion using the explicit model Hamiltonian. Here we adopt a solvable model~\cite{Lee:1954iq}, which consists of two static fermions $B_{0}$ and $N$ and a scalar boson $\theta$.\footnote{We rename the original $V$ particle as $B_{0}$, in order to be consistent with the notations in the previous sections. The statistical nature of the particles is specified simply for definiteness of the discussion.} 

\begin{figure}[bt]
\centerline{
\psfig{file=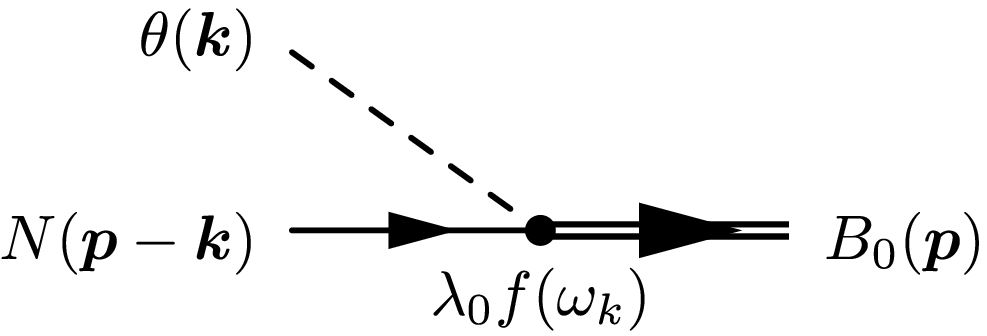,width=4.7cm}
\hspace{1cm}
\psfig{file=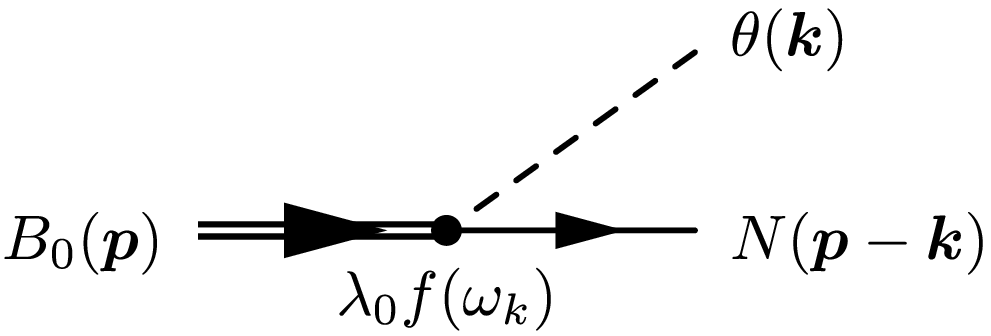,width=4.7cm}}
\vspace*{8pt}
\caption{Diagrammatic representations of the interaction Hamiltonian~\eqref{eq:Hint}. \label{fig:int}}
\end{figure}

The Hamiltonian of the model consists of the free part and the interaction term,
\begin{align}
    H
    & =H_{0}+V\;  .
\end{align}
The free Hamiltonian is given by
\begin{align}
    H_{0}
    & =m_{B_{0}}\int d\bm{p}
    B_{0}^{\dag}(\bm{p})B_{0}(\bm{p})
    +m_{N}\int d\bm{q}N^{\dag}(\bm{q})N(\bm{q})
    +\int d\bm{k} \omega_{k}
    \theta^{\dag}(\bm{k})\theta(\bm{k})\;  ,
\end{align}
with $\omega_{k}=\sqrt{\bm{k}^{2}+m_{\theta}^{2}}$, $m_{N}$ and $m_{\theta}$ being the mass of the $N$ and $\theta$ fields, and $m_{B_{0}}$ being the bare mass of the $B_{0}$ field. The bare binding energy in the previous sections corresponds to
\begin{align}
    B_{0}
    & =
    m_{N}+m_{\theta}-m_{B_{0}}\;  ,
\end{align}
and we assume that $B_{0}>0$. The interaction Hamiltonian is chosen to be
\begin{align}
    V
    & =
    \frac{\lambda_{0}}{(2\pi)^{3/2}}
    \int d\bm{k}d\bm{p}
    \frac{f(\omega_{k})}{\sqrt{2\omega_{k}}}
    \{B_{0}^{\dag}(\bm{p})N(\bm{p}-\bm{k})\theta(\bm{k})
    +N^{\dag}(\bm{p}-\bm{k})B_{0}(\bm{p})\theta^{\dag}(\bm{k})
    \}\;  ,
    \label{eq:Hint}
\end{align}
where $\lambda_{0}$ is the bare coupling constant and $f(\omega_{k})$ represents the energy dependence of the coupling. The first (second) term annihilates the $\theta N$ state (the $B_{0}$ particle) and creates the $B_{0}$ particle (the $\theta N$ state). These are diagrammatically represented in Fig.~\ref{fig:int}. The interaction conserves the fermion number $n_{B_{0}}+n_{N}$ as well as the sum of the numbers of the $\theta$ and $B_{0}$ particles $n_{\theta}+n_{B_{0}}$, where $n_{i}$ is the total number of the particle $i$.

The vacuum of the model $\ket{0}$ is specified by $B_{0}(\bm{p})\ket{0}=N(\bm{p})\ket{0}=\theta(\bm{p})\ket{0}=0$. We construct the $N\theta$ scattering state and the bare bound state as
\begin{align}
    \ket{\bm{p}}
    & =
    N^{\dag}(\bm{0})\theta^{\dag}(\bm{p})\ket{0}\; , 
    \label{eq:modelp}\\
    \ket{B_{0}}
    &= B_{0}^{\dag}(\bm{0})\ket{0}\;  ,
\end{align}
in the rest frame of the fermions. These are the eigenstates of the free Hamiltonian:
\begin{align}
    H_{0}\ket{\bm{p}}
    & =
    (m_{N}+\omega_{p})\ket{\bm{p}}\;  , \\
    H_{0}\ket{B_{0}}
    &= m_{B_{0}}\ket{B_{0}}\;  .
\end{align}
In the $n_{B_{0}}+n_{N}=n_{\theta}+n_{B_{0}}=1$ sector, $\ket{B_{0}}$ and $\ket{\bm{p}}$ are the only eigenstates. It is shown that the states $N^{\dag}(\bm{0})\ket{0}$ and $\theta^{\dag}(\bm{p})\ket{0}$ are also the eigenstates of the full Hamiltonian, so we have 
\begin{align}
    H\ket{\bm{p}}
    & =
    (m_{N}+\omega_{p})\ket{\bm{p}}\;  .
\end{align}
In other words, the $N$ and $\theta$ fields are not renormalized by the interaction~\eqref{eq:Hint}. On the other hand, $\ket{B_{0}}$ is not the eigenstate of the full Hamiltonian, because we have
\begin{align}
    V\ket{B_{0}}
    & =
    \frac{\lambda_{0}}{(2\pi)^{3/2}}\int d\bm{p}
    \frac{f(\omega_{p})}{\sqrt{2\omega_{p}}}
    \ket{\bm{p}}\;  .
    \label{eq:modelV}
\end{align}
Thus, we need to renormalize the $B_{0}$ field. We define the renormalized field $\ket{B}$ so as to satisfy the eigenvalue equation 
\begin{align}
    H\ket{B}
    & =
    m_{B}\ket{B}\;  .
    \label{eq:eigenvalue}
\end{align}
Because $\ket{B_{0}}$ and $\ket{\bm{p}}$ spans the basis in the $n_{B_{0}}+n_{N}=n_{\theta}+n_{B_{0}}=1$ sector, the field $\ket{B}$ can be expressed by the linear combination of these states. We denote it as
\begin{align}
    \ket{B}
    & = 
    \sqrt{Z}
    \left[
    \ket{B_{0}}
    +
    \int d\bm{p} \; \Phi(\bm{p})\ket{\bm{p}}
    \right] \; ,
    \label{eq:ketB}
\end{align}
where the constant $Z$ and the function $\Phi(\bm{p})$ are to be determined. It is clear that the field renormalization constant $Z$ is defined in the same way with Eq.~\eqref{eq:Zdef} because 
\begin{align}
    |\bra{B_{0}}\kket{B}|^{2}
    & = 
    Z\;  .
\end{align}
Therefore, $Z$ represents the probability of finding the bare state $\ket{B_{0}}$ in the physical state $\ket{B}$. Multiplying $\bra{\bm{p}}$ to Eq.~\eqref{eq:eigenvalue}, we determine the function $\Phi(\bm{p})$ as
\begin{align}
    \frac{\lambda_{0}}{(2\pi)^{3/2}}
    \frac{f(\omega_{p})}{\sqrt{2\omega_{p}}}
    +\Phi(\bm{p})(m_{N}+\omega_{p})
    & = 
    m_{B}\Phi(\bm{p})\; , \\
    \Phi(\bm{p})
    & = 
    \frac{\lambda_{0}}{(2\pi)^{3/2}}
    \frac{f(\omega_{p})}{\sqrt{2\omega_{p}}}
    \frac{1}{m_{B}-m_{N}-\omega_{p}}\;  .
    \label{eq:Phi}
\end{align}
By the normalization $\bra{B}\kket{B}=1$, we obtain 
\begin{align}
    1
    & 
    =Z\left[
    1
    +
    \int d\bm{p} |\Phi(\bm{p})|^{2}
    \right] \label{eq:1decomp}\; .
\end{align}
Substituting Eq.~\eqref{eq:Phi} into Eq.~\eqref{eq:1decomp}, we arrive at the expression
\begin{align}
    Z
    & 
    =1-\frac{\lambda^{2}}{(2\pi)^{3}}
    \int d\bm{p}
    \frac{|f(\omega_{p})|^{2}}{2\omega_{p}}
    \frac{1}{(m_{B}-m_{N}-\omega_{p})^{2}}\;  ,
    \label{eq:ZLee}
\end{align}
where the renormalized coupling constant is defined as $\lambda^{2}=\lambda_{0}^{2}Z$. This is equivalent to Eq.~\eqref{eq:Xgeneral}, because it follows from Eqs.~\eqref{eq:modelp}, \eqref{eq:modelV} and \eqref{eq:ketB} that
\begin{align}
    \bra{\bm{p}}V\ket{B}
    & 
    =
    \frac{\lambda}{(2\pi)^{3/2}}\frac{f(\omega_{p})}{\sqrt{2\omega_{p}}}
\end{align}
and $\omega_{p}-m_{\theta} \approx E_{p}$ for small $|\bm{p}|$. 

\subsection{Relation to the scattering amplitude}\label{subsec:amp}

Equation~\eqref{eq:Xexact} expresses the compositeness $X$ by the form factor $G(E)$. The compositeness $X$ can also be expressed by the two-body scattering amplitude. The $T$-operator $T(z)$ is defined to satisfy the Lippmann-Schwinger equation\cite{Taylor}
\begin{align}
    T(z)
    &=V+V\frac{1}{z-H_{0}}T(z)  
    \; ,
    \label{eq:LS}
\end{align}
for a complex energy variable $z$. From the $T$-operator, the on-shell $t$-matrix is obtained as
\begin{align}
    t(E_{p}) &= \bra{\bm{p}^{\prime}}T(E_{p}+ i0)\ket{\bm{p}}\;  ,
    \quad E_{p^{\prime}}=E_{p}
    \label{eq:onshellt}\;  ,
\end{align}
which depends only on the energy $E_{p}$ for the $s$-wave scattering. 

The formal solution of Eq.~\eqref{eq:LS} is given by
\begin{align}
    T(z)
    &=V+V\frac{1}{z-H}V  
    \; . 
\end{align}
Inserting the complete set of the in state shown in Eq.~\eqref{eq:completenessfull}, we obtain
\begin{align}
    T(z)
    &=V
    +\frac{V\ket{B}\bra{B}V}{z+B} 
    +\int d\bm{q}
    \frac{V\ket{\bm{q},+}\bra{\bm{q},+}V}{z-E_{q}}
    \; . 
\end{align}
Now, we use the relation
\begin{align}
    T(E_{p}\pm i0)\ket{\bm{p}} &= V\ket{\bm{p},\pm}\;  , \\
    \bra{\bm{p}}T(E_{p}\pm i0) &= \bra{\bm{p},\mp}V\;   ,
\end{align}
which leads to
\begin{align}
    T(z)
    &=V
    +\frac{V\ket{B}\bra{B}V}{z+B} 
    +\int d\bm{q}
    \frac{T(E_{q}+ i0)\ket{\bm{q}}\bra{\bm{q}}T(E_{q}- i0)}{z-E_{q}} \\
    &=V
    +\frac{V\ket{B}\bra{B}V}{z+B} 
    +\int d\bm{q}
    \frac{T(E_{q}+ i0)\ket{\bm{q}}\bra{\bm{q}}T^{*}(E_{q}+ i0)}{z-E_{q}}
    \; ,
\end{align}
where we have used $T(z^{*})=T^{*}(z)\equiv [T(z)]^{*}$. Taking the matrix element of this operator by $\bra{\bm{p}^{\prime}}$ and $\ket{\bm{p}}$ and setting $z=E_{p}+i0$, we obtain 
\begin{align}
    \bra{\bm{p}^{\prime}}T(E_{p}+i0)\ket{\bm{p}}
    &=\bra{\bm{p}^{\prime}}V\ket{\bm{p}}
    +\frac{\bra{\bm{p}^{\prime}}V\ket{B}\bra{B}V\ket{\bm{p}}}{E_{p}+i0+B} 
    \nonumber \\
    &\quad +\int d\bm{q}
    \frac{\bra{\bm{p}^{\prime}}T(E_{q}+ i0)\ket{\bm{q}}
    \bra{\bm{q}}T^{*}(E_{q}+ i0)\ket{\bm{p}}}{E_{p}+i0-E_{q}} 
    \; ,
    \nonumber \\
    t(E_{p})
    &=v
    +\frac{|G(E_{p})|^2}{E_{p}+B} 
    +\int d\bm{q}
    \frac{|t(E_{q})|^{2}}{E_{p}-E_{q}+i0}\;  ,
\end{align}
where $v\equiv \bra{\bm{p}^{\prime}}V\ket{\bm{p}}$ and we have used Eqs.~\eqref{eq:FF}, \eqref{eq:onshellt}, and the condition $E_{p^{\prime}}=E_{p}$. We now obtain the Low's equation by rewriting the variables as
\begin{align}
    t(E)
    &=v
    +
    \frac{|G(E)|^2}{E+B} 
    +4\pi\sqrt{2\mu^3}\int_0^{\infty} dE^{\prime}
    \frac{\sqrt{E^{\prime}}|t(E^{\prime})|^2}
    {E-E^{\prime}+i0}\;  .
    \label{eq:Low}
\end{align}
Because the second term includes the form factor $|G(E)|^{2}$ in the integrand of \eqref{eq:Xexact}, the compositeness can be written by the scattering amplitude~\cite{Hyodo:2010uh,Hyodo:2011qc}
\begin{align}
    X
    &=4\pi\sqrt{2\mu^3}
    \int_{0}^{\infty} dE \frac{\sqrt{E}}{E+B}
    \Biggl[
    t(E)-v 
    -
    4\pi\sqrt{2\mu^3}\int_{0}^{\infty} dE^{\prime}
    \frac{\sqrt{E^{\prime}}|t(E^{\prime})|^2}{E-E^{\prime}+i0}
    \Biggr]\;  .
    \label{eq:Xexactamp}
\end{align}
By definition, the compositeness $X$ is a real number. On the other hand, the scattering amplitude $t(E)$ in the integrand is complex for $E> 0$. The imaginary part of $t(E)$ is given by the optical theorem as
\begin{align}
    \im t(E)
    &=-4\pi^{2} \mu q(E)|t(E)|^2
    =-4\pi^{2} \sqrt{ 2\mu^{3}E}|t(E)|^2
    \label{eq:optical}\;  ,
\end{align}
where $q(E)=\sqrt{2\mu E}$. This imaginary part is exactly cancelled by the imaginary part of the integral of the last term in Eq.~\eqref{eq:Xexactamp} as
\begin{align}
    \text{Im} \Biggl[
    -
    &4\pi\sqrt{2\mu^3}\int_{0}^{\infty} dE^{\prime}
    \frac{\sqrt{E^{\prime}}|t(E^{\prime})|^2}{E-E^{\prime}+i0}
    \Biggr] \nonumber \\
    &= 
    -4\pi\sqrt{2\mu^3} (-\pi)\sqrt{E}|t(E)|^2
    =-\im t(E) \; .
\end{align} 
To express this cancellation explicitly, we can rewrite the compositeness as
\begin{align}
    X
    &=4\pi\sqrt{2\mu^3}
    \int_{0}^{\infty} dE \frac{\sqrt{E}}{E+B}
    \Biggl[
    \re t(E)-v 
    -
    4\pi\sqrt{2\mu^3}\mathcal{P}\int_{0}^{\infty} dE^{\prime}
    \frac{\sqrt{E^{\prime}}|t(E^{\prime})|^2}{E-E^{\prime}}
    \Biggr]\; ,
\end{align}
where $\mathcal{P}$ stands for the principal value integration. In this way, the compositeness can be expressed by the scattering amplitude $t(E)$.

\subsection{Weak-binding limit}\label{subsec:weakbinding}

So far we have not introduced any approximations. The compositeness is given either by the form factor $G(E)$ or by the scattering amplitude $t(E)$. In both cases, the knowledge of the nonperturbative solution of the two-body problem (the wave function of the bound state) is necessary to determine the compositeness $X$. 

It is however shown\cite{Weinberg:1965zz} that the compositeness is model-independently determined in the weak-binding limit. We now consider the case where the binding energy is small in comparison with the typical energy scale of the interaction $E_{\rm typ}$,
\begin{align}
    B
    & \ll E_{\rm typ}\; .
\end{align}
In this case, the integration of Eq.~\eqref{eq:Xexact} is dominated by the energy region $E \lesssim E_{\rm typ}$, because the $1/(E+B)^{2}$ factor enhances the small $E$ region. We, therefore, expand the transition form factor $|G(E)|^{2}$ around $E=0$ as
\begin{align}
    |G(E)|^2
    &=g_{0}^{2}
    +Eg_{1}^{2}
    +\cdots , \label{eq:Gexpand}\\
    g_{0}^{2}
    &\equiv |G(0)|^2, \quad
    g_{1}^{2}
    \equiv 
    \left.\frac{d|G(E)|^2}{dE}\right|_{E=0}\; ,
\end{align}
where $g_{0}$ is regarded as the coupling constant. The compositeness $X$ is then given by
\begin{align}
    X
    &\approx
    4\pi\sqrt{2\mu^3}
    \int_{0}^{E_{\rm typ}} dE\frac{\sqrt{E}
    |G(E)|^2}{(E+B)^2} \quad (B\ll E_{\rm typ})\nonumber \\
    &=
    4\pi\sqrt{2\mu^3}
    g_{0}^{2}
    \int_{0}^{E_{\rm typ}} dE\frac{\sqrt{E}}{(E+B)^2} 
    +4\pi\sqrt{2\mu^3}g_{1}^{2}
    \int_{0}^{E_{\rm typ}} dE\frac{\sqrt{E}E}{(E+B)^2} 
    +\cdots \nonumber \\
    &=
    4\pi\sqrt{2\mu^3}
    g_{0}^{2}
    \left[
    \frac{1}{\sqrt{B}}\arctan{\sqrt{\frac{E_{\rm typ}}{B}}}
    -\frac{\sqrt{E_{\rm typ}}}{B+E_{\rm typ}}
    \right] \nonumber \\
    &\quad +4\pi\sqrt{2\mu^3}g_{1}^{2}
    \left[
    -3\sqrt{B}\arctan{\sqrt{\frac{E_{\rm typ}}{B}}}
    +\frac{B\sqrt{E_{\rm typ}}}{B+E_{\rm typ}}
    +2\sqrt{E_{\rm typ}}
    \right]
    +\cdots \nonumber \\
    &=
    4\pi\sqrt{2\mu^3}
    \left[
    \frac{g_{0}^{2}}{\sqrt{B}}\left(\frac{\pi}{2}
    +\mathcal{O}(\sqrt{B/E_{\rm typ}})\right)
    +\frac{g_{0}^{2}}{\sqrt{E_{\rm typ}}}\cdot \mathcal{O}(1)
    +g_{1}^{2}\sqrt{E_{\rm typ}}\cdot \mathcal{O}(1)
    \right] +\cdots .\nonumber 
\end{align}
We assume that the form factor is well-behaved around $E=0$ so that the energy derivative of the form factor can be estimated by the typical energy scale $E_{\rm typ}$ as 
\begin{align}
    g_{1}^{2}
    &\sim
    \frac{g_{0}^{2}}{E_{\rm typ}}\; .
\end{align}
It is also reasonable to assume $g_{n}\sim g_{n-1}^{2}/E_{\rm typ}$ in the higher-order derivatives. We then obtain the leading order result of the $B/E_{\rm typ}$ expansion of the compositeness $X$ as
\begin{align}
    X
    &=1-Z\approx
    2\pi^{2}\sqrt{2\mu^3}
    \frac{g_{0}^{2}}{\sqrt{B}}
     \quad (B\ll E_{\rm typ})
    \label{eq:XwB}\; .
\end{align}
This is the first main result in the weak-binding limit; the compositeness $X$ is determined by the binding energy $B$ and the coupling constant $g_{0}=G(E=0)$ instead of the form factor function $G(E)$. Since $X\leq 1$, the upper limit of the coupling strength can be obtained as 
\begin{align}
    g_0^2
    &\leq 
    \frac{1}{2\pi^2}  
    \sqrt{\frac{B}{2\mu^3}}\;  .
\end{align}
For a purely composite particle, we have $X=1$ and the equality holds. This is the Weinberg's compositeness condition for the coupling constant. It is instructive to derive Eq.~\eqref{eq:XwB} from the scattering amplitude~\eqref{eq:Xexactamp}. For a small $B$, the bracket in Eq.~\eqref{eq:Xexactamp} is dominated by the bound state pole term in the scattering amplitude $t(E)$ as
\begin{align}
    X
    & \approx 4\pi\sqrt{2\mu^3}
    \int_{0}^{\infty} dE \frac{\sqrt{E}}{E+B}
    \cdot
    \left(
    \frac{g^{2}}{E+B}
    \right) 
    \quad (B\ll E_{\rm typ})\; ,
\end{align}
where $g\equiv G(E=-B)$. Noting that $g^{2}=g_{0}^{2}+\mathcal{O}(B/E_{\rm typ})$, we obtain Eq.~\eqref{eq:XwB}.

Next, we take the weak-binding limit of the Low's equation~\eqref{eq:Low}. In order to concentrate on the low-energy behavior of the amplitude near the threshold, we consider the energy region $E\sim B$. In this case, we are left with the first term of the expansion of the form factor~\eqref{eq:Gexpand} as $|G(E)|^{2}\approx g_{0}^{2}$. Because the compositeness $X$ is of the order of $\mathcal{O}(1)$, Eq.~\eqref{eq:XwB} indicates that $g_{0}^{2}\sim \mathcal{O}(\sqrt{B/E_{\rm typ}})$. This means that, the second term of Eq.~\eqref{eq:Low} $g_{0}^{2}/(E+B)$ is of order $\mathcal{O}(\sqrt{E_{\rm typ}/B})$, which should be much larger than the ($B$-independent) interaction $v\sim \mathcal{O}(1)$. Then we neglect $v$ in Eq.~\eqref{eq:Low} and obtain the equation
\begin{align}
    t(E)
    &\approx 
    \frac{g_{0}^{2}}{E+B} 
    +4\pi\sqrt{2\mu^3}\int_0^{\infty} dE^{\prime}
    \frac{\sqrt{E^{\prime}}|t(E^{\prime})|^2}
    {E-E^{\prime}+i0} 
    \quad (E\sim B\ll E_{\rm typ})
    \label{eq:Lowapprox}\; .
\end{align}
This integral equation can be solved by considering the analytic property of the function
\begin{align}
    \tau(z)
    &= 
    \frac{g_{0}^{2}}{z+B} 
    +4\pi\sqrt{2\mu^3}\int_0^{\infty} dE^{\prime}
    \frac{\sqrt{E^{\prime}}|t(E^{\prime})|^2}
    {z-E^{\prime}} \; ,
\end{align}
with the complex energy variable $z$. The solution is\cite{Weinberg:1965zz}
\begin{align}
    t(E)
    &=\left[
    \frac{E+B}{g_{0}^{2}}
    +\frac{4\pi^{2}\sqrt{2\mu^3} (B-E)}{2\sqrt{B}}
    +i4\pi^{2}\sqrt{2\mu^3} \sqrt{E}
    \right]^{-1} .
    \label{eq:tmatrix}
\end{align}
In the form of the scattering amplitude $f(p)=-4\pi^{2}\mu t(E_{p})$, we have
\begin{align}
    f(p)
    &=\left[
    -\frac{B}{4\pi^{2}\mu g_{0}^{2}}
    -\frac{\sqrt{2\mu B}}{2}
    -ip
    +
    \frac{1}{2}
    \left(
    -\frac{1}{4\pi^{2}\mu^{2}g_{0}^{2}}
    +\frac{1}{\sqrt{2\mu B}}
    \right)
    p^{2}
    \right]^{-1} .
    \label{eq:famplitude}
\end{align}
This is exactly the same functional form with the effective range expansion truncated up to $p^{2}$ order,
\begin{align}
    f(p)
    &=\left[
    -\frac{1}{a}
    -ip
    +\frac{r_{e}}{2}p^{2}
    \right]^{-1} ,
    \label{eq:effectiverangeExp}
\end{align}
where $a$ is the scattering length and $r_{e}$ is the effective range. It should be noted that we have restricted ourselves to the low-energy region $E\sim B$ in the derivation of Eq.~\eqref{eq:Lowapprox}, so the result~\eqref{eq:famplitude} is valid only near the threshold. At higher energies, the neglected contributions in Eq.~\eqref{eq:Lowapprox} will generate the higher-order terms in the effective range expansion. Comparing Eqs.~\eqref{eq:famplitude} and \eqref{eq:effectiverangeExp}, we determine the scattering length and the effective range as
\begin{align}
    a
    &=2R\left(
    1+\frac{\sqrt{B}}{2\pi^{2}\sqrt{2\mu^{3}}g_{0}^{2}}
    \right)^{-1} , \\
    r_{e}
    &=R\left(
    1-\frac{\sqrt{B}}{2\pi^{2}\sqrt{2\mu^{3}}g_{0}^{2}}
    \right) ,\\
    R&\equiv
    \frac{1}{\sqrt{2\mu B}}\; .
\end{align}
This is the second main result in the weak-binding limit; the scattering length $a$ and the effective range $r_{e}$ are related to the binding energy $B$ and the coupling constant $g_{0}$. By using the first result Eq.~\eqref{eq:XwB}, we can eliminate the coupling constant $g_{0}^{2}$. In this case, we obtain
\begin{align}
    a
    =\frac{2X}{X+1}R + \mathcal{O}(R_{\rm typ})
    &=\frac{2(1-Z)}{2-Z}R + \mathcal{O}(R_{\rm typ})\; ,
    \label{eq:scatteringlength} \\
    r_{e}
    =\frac{X-1}{X}R+ \mathcal{O}(R_{\rm typ})
    &=\frac{-Z}{1-Z}R+ \mathcal{O}(R_{\rm typ})\; , 
    \label{eq:effectiverange} \\
    R_{\rm typ}
    &\equiv \frac{1}{\sqrt{2\mu E_{\rm typ}}}\; ,
\end{align}
where $R_{\rm typ}$ is the typical length scale of the potential. Because we have neglected the $\mathcal{O}(1)$ contributions in comparison with $\mathcal{O}(\sqrt{E_{\rm typ}/B})$, the uncertainty of the approximation is estimated as $\mathcal{O}(R_{\rm typ})$. These are the final results of Ref.~\citen{Weinberg:1965zz}.

Let us summarize the discussion in the weak-binding. We first derive the exact expression of the compositeness $X=1-Z$ as Eqs.~\eqref{eq:Xexact} and \eqref{eq:Xexactamp}. These requires the solution of the Schr\"odinger equation $\ket{B}$ and the solution of the Lippmann--Schwinger equation $t(E)$, respectively. This is possible when we specify the explicit form of the Hamiltonian $H$, or equivalently, the explicit form of the potential $V$. Thus, the compositeness of a general bound state depends on the choice of the potential.

We then consider the weak-binding limit $B\ll E_{\rm typ}$, to obtain the expressions~\eqref{eq:XwB}, \eqref{eq:scatteringlength} and \eqref{eq:effectiverange}. In this case, the explicit dependence on $V$ and $\ket{B}$ is lost in Eq.~\eqref{eq:XwB} as $\bra{\bm{p}}V\ket{B}=G(E)\to g_{0}$. In the same way, $v=\bra{\bm{p}^{\prime}}V\ket{\bm{p}}$ is dropped in Eq.~\eqref{eq:Lowapprox}. The effect of the potential $V$ is then exclusively included in the coupling constant $g_{0}$, scattering length $a$, and the effective range $r_{e}$. Because $a$ and $r_{e}$ are the experimental observables, the final results~\eqref{eq:scatteringlength} and \eqref{eq:effectiverange} are \textit{independent} of the potential $V$. In this sense, this is a kind of the universality of the structure of the weakly bound state. The property of the bound state does not depend on the detailed dynamics of the interaction, and completely specified by the values of a few threshold parameters in the weak-binding limit.

We have emphasized that the result in the weak-binding limit is model-independent, while the exact expression of the compositeness is not. Formally, the ambiguity comes from the choice of the bare Hamiltonian $H_{0}$ in Eq.~\eqref{eq:decomposition}. This is equivalent to the ambiguity of the choice of the potential $V$. Even though the full Hamiltonian $H$ is given, the choice of the basis (eigenstates of $H_{0}$) to decompose the bound state wave function is not unique. With a different basis, the value of the compositeness will change. In this sense, this is similar to the nonuniqueness of the hadron potential in lattice QCD.~\cite{Aoki:2012tk} We need additional criterion to choose a ``good'' basis for the discussion of the compositeness.

\subsection{Interpretation of the elementary contribution}\label{subsec:interpretation}

Let us consider the implication of Eqs.~\eqref{eq:scatteringlength} and \eqref{eq:effectiverange}. If the bound state is purely elementary $X=0$ ($Z=1$), then we have
\begin{align}
    a
    &=0,\quad
    r_{e}=-\infty
    \quad \text{(purely elementary limit)}\; ,
    \label{eq:pureelementary}
\end{align}
where we have neglected the error term of $\mathcal{O}(R_{\rm typ})$. In contrast, if the bound state is purely composite $X=1$ ($Z=0$), then we have
\begin{align}
    a
    &=R,\quad
    r_{e}=0 
    \quad \text{(purely composite limit)}\; .
    \label{eq:purecomposite}
\end{align}
Thus, including the error term, we find the criterion of the structure of the bound states by the scattering length and the effective range as
\begin{align}
    \begin{cases}
    a\sim R_{\rm typ}
    \ll -r_{e} & \text{(elementary dominance)}\; ,  \\
    a\sim R
    \gg r_{e}\sim R_{\rm typ} & \text{(composite dominance)}\; .
    \end{cases}\label{eq:criterion}
\end{align}

We notice that Eq.~\eqref{eq:effectiverange} always gives a negative effective range for $0\leq Z\leq 1$, up to the correction of $\mathcal{O}(R_{\rm typ})$. This is in contradiction to the intuitive picture in which the effective range represents the mean distance of the interaction and hence should be positive~\cite{BlattWeisskopf}. However, this interpretation of the effective range is ensured only for simple attractive potentials without energy dependence. It is shown that the effective range can be negative in the renormalizable local Hamiltonian quantum field theory with derivative interactions\cite{Phillips:1997xu,Braaten:2007nq}. Because the momentum dependence of the interaction can be translated into the energy dependence which can be further interpreted as a consequence of the elimination of the coupled-channel effect~\cite{Feshbach:1958nx,Feshbach:1962ut}, a large and negative $r_{e}$ represents the effect of the contributions other than the scattering state of interest (recall the schematic discussion in Subsec.~\ref{subsec:bare}). In this sense, the elementary contribution comes from the outside of the two-body scattering model space, so it can be identified as the CDD pole contribution. 

It is also instructive to discuss the purely composite case $X=1$ ($Z=0$) in more detail.\footnote{The author thanks Koichi Yazaki for the useful discussion on this point.} Let us consider a simple square well potential without the energy- and momentum-dependence. Since this potential  corresponds to the energy-independent interaction in the field theoretical models, we always have $Z=0$ and the generated bound state is a purely composite state.\cite{PR136.B816,Hyodo:2011qc} On the other hand, it is possible to produce arbitrary threshold parameters by adjusting the depth and range of the potential well. Is this consistent with the criterion~\eqref{eq:criterion}?

Here, we should remember the weak-binding assumption; the binding energy should be sufficiently small for the criterion \eqref{eq:criterion} to work. As mentioned above, the simple attractive potential produces a positive effective range of the order of $R_{\rm typ}$. Thus, it is not likely that the elementary dominance of Eq.~\eqref{eq:criterion} is satisfied. On the other hand, when the binding energy is small, $R=1/\sqrt{2\mu B}$ gets large. When $R$ is sufficiently larger than $r_{e}\sim R_{\rm typ}$, the amplitude~\eqref{eq:effectiverangeExp} is approximated by $f(p)\sim (-1/a-ip)^{-1}$ whose pole condition is given by $a\sim R$. Thus, eventually the composite dominance is satisfied. Although the simple attractive potential can produce arbitrary threshold parameters, only those with small binding energy can be judged by the criterion~\eqref{eq:criterion} and it indeed deduces the dominance of the composite component.

In this way, the ``composite dominance'' in Eq.~\eqref{eq:criterion} is understood as a natural consequence of the simple attractive potential in the weak-binding limit. If there is nothing special in the potential, we always have $a\sim R$ for a sufficiently small binding energy. In contrast, some additional contribution other than the two-body state leads to the negative $r_{e}$. The ``elementary dominance'' is understood that the large CDD pole contribution makes the effective range anomalously large in the negative direction.

\subsection{Application to the deuteron}\label{subsec:deuteron}

Let us apply the above argument to the deuteron, the bound state of two nucleons in the $^{3}S_{1}$ channel. We regard the physical deuteron as the bound state $\ket{B}$ and the two-nucleon $NN$ states as the scattering states. Bearing the pion exchange force in mind, we estimate the typical (smallest) momentum scale of the $NN$ interaction as  $k_{\rm typ}\sim m_{\pi}$. The typical energy scale is then given by
\begin{align}
    E_{\rm typ}
    & =\frac{k_{\rm typ}^{2}}{2\mu}
    \sim \frac{m_{\pi}^{2}}{M_{N}}\sim 20 \text{ MeV}\;  .
\end{align}
The deuteron binding energy $B\sim 2$ MeV is an order of magnitude smaller than this scale $E_{\rm typ}$, so we can apply the weak-binding argument.

The experimental values of the threshold parameters of the $NN$ scattering in the $^{3}S_{1}$ channel and the deuteron radius are~\cite{Machleidt:2000ge}
\begin{align}
    a
    &= 5.419\pm 0.007 \text{ fm}\; , \\
    r_{e}
    &= 1.7513\pm 0.008 \text{ fm}\; , \\
    R
    &= 4.31767 \text{ fm}\; .
\end{align}
where we have used $B= 2.224575$ MeV to estimate $R$. With $R_{\rm{typ}}\sim 1/m_{\pi}\sim 1.43$ fm, we find that the deuteron case is judged as the composite dominance in Eq.~\eqref{eq:criterion}, and we conclude that the deuteron is dominated by the two-nucleon composite component. 

We should keep in mind that this conclusion is drawn by the leading-order result in the weak-binding expansion. In reality, the deuteron structure is more complicated, for instance, by the existence of the $d$-wave component~\cite{Machleidt:2000ge}. Nevertheless, we again emphasize that the ``composite dominance'' of the deuteron is derived without using the nuclear force potential and the deuteron wave function. As a matter of fact, the $NN$ scattering is a fortunate channel in the sense that a large number of experimental data with high accuracy enables us to construct the realistic nuclear forces.\cite{Wiringa:1994wb,Machleidt:2000ge} Except for the $NN$ scattering, we do not have realistic potentials of other hadron--hadron interaction, simply because of the lack of the experimental data. Thus, the structure of hadrons cannot be extracted from their reliable wave functions. Equation~\eqref{eq:criterion} tells us that the structure of the weakly bound state can still be determined even in these cases, with the knowledge of the scattering length and the effective range.

\subsection{Generalization to the multichannel case}\label{subsec:multi}

\begin{figure}[tpb]
\centerline{
\psfig{file=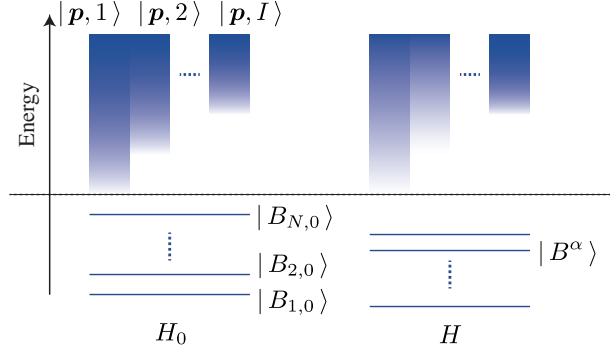,width=8cm}
}
\vspace*{8pt}
\caption{Schematic illustration of the spectra of the free Hamiltonian $H_{0}$ (left) and the full Hamiltonian $H$ (right) in the multichannel scattering. \label{fig:spectrum2}}
\end{figure}

Here, we present the formulation with multiple bare bound stats and scattering states.\cite{Baru:2003qq,Hyodo:2011qc} The decomposition of the Hamiltonian is same as before,
\begin{align}
    H
    & = H_{0}+V\;  .
\end{align}
Now, we include $N$ bare bound states labeled by the index $n$, and $I$ scattering channels by the index $i$ (see Fig.~\ref{fig:spectrum2}). The eigenstates of the free Hamiltonian are summarized as
\begin{align}
    H_0\ket{\bm{p},i}
    & = E_{p,i}\ket{\bm{p},i} 
    \quad
    (i=1,...,I) \; , \\
    E_{p,i}
    &\equiv 
    \frac{\bm{p}^{2}}{2\mu_{i}}+E_{\text{th},i} \; , \\
    H_0\ket{B_{n,0}}
    & = -B_{n,0}\ket{B_{n,0}}  
    \quad
    (n=1,...,N) \; , 
\end{align}
where $\mu_{i}$ is the reduced mass of the scattering channel $i$, $E_{\text{th},i}$ is the threshold energy difference from the reference channel $i=1$. We define $E_{\text{th},1}=0$ and $E_{\text{th},i}\geq E_{\text{th},j}$ for $i>j$ without loss of generality. The labels $i$ and $n$ can also be used for the internal degrees of freedom. In this case, there is a degeneracy in $E_{\text{th},i}$ and $B_{n,0}$, respectively. The eigenstates are orthogonal to each other 
\begin{align}
    \bra{\bm{p},i}\kket{\bm{p}^{\prime},j}
    &=\delta(\bm{p}^{\prime}-\bm{p})\delta_{ij}\; , 
    \quad \bra{B_{n,0}}\kket{B_{m,0}}
    =\delta_{nm}\; , 
    \quad \bra{B_{n,0}}\kket{\bm{p},i}
    =0 \;  .
\end{align}
The completeness relation is given by
\begin{align}
    1 
    &= \sum_{n=1}^{N}\ket{B_{n,0}}\bra{B_{n,0}} 
    + \sum_{i=1}^{I}\int d\bm{p} \ket{\bm{p},i}\bra{\bm{p},i}\;  .
    \label{eq:completenessmulti}
\end{align}

We use this basis to decompose the physical bound state $\ket{B^{\alpha}}$, which is an eigenstate of the full Hamiltonian
\begin{align}
    H\ket{B^{\alpha}}
    &= -B^{\alpha}\ket{B^{\alpha}}\;  ,
\end{align}
with $B^{\alpha}>0$. For the discussion of the structure of the bound state $\ket{B^{\alpha}}$, it is not necessary to specify the basis of the full Hamiltonian, but we need to require the normalization of the physical state
\begin{align}
    \bra{B^{\alpha}}\kket{B^{\alpha}}
    &= 1
    \label{eq:normalizationmulti}\;  .
\end{align}
We then define $Z_{n}^{\alpha}$ and $X_i^{\alpha}$ as 
\begin{align}
    Z_{n}^{\alpha}
    &=|\bra{B_{n,0}}\kket{B^{\alpha}}|^2\; , \\
    X_i^{\alpha}
    &=\int d\bm{p}|\bra{\bm{p},i}\kket{B^{\alpha}}|^2\; , 
\end{align}
where $Z_{n}^{\alpha}$ ($X_i^{\alpha}$) represents the probability of finding the bound state $\ket{B^{\alpha}}$ in the bare state $\ket{B_{n,0}}$ (in the scattering state $\ket{\bm{p},i}$). Thus, we regard $Z_{n}^{\alpha}$ ($X_i^{\alpha}$) as the elementariness of the $n$th bare state (compositeness of the scattering state in the channel $i$). 
We can also define the whole elementariness and whole compositeness as sums of each component,
\begin{align}
    Z^{\alpha}
    &= \sum_{n=1}^{N}Z_{n}^{\alpha}\;  ,\\
    X^{\alpha}
    &= \sum_{i=1}^{I} X_i^{\alpha}\;  ,
\end{align}
where $Z^{\alpha} (X^{\alpha})$ represents the probability of finding the bound state $\ket{B^{\alpha}}$ in any one of the bare states (the scattering states). Thanks to the completeness relation~\eqref{eq:completenessmulti} and the normalization~\eqref{eq:normalizationmulti}, we find 
\begin{align}
    1 
    &= Z^{\alpha} + X^{\alpha} 
    \label{eq:normalizationmulti2}
\end{align}
and its probability interpretation is guaranteed.

Following the same argument with section~\ref{subsec:Zdef}, we obtain
\begin{align}
    X^{\alpha}
    &= 
    \sum_{i}
    \int d\bm{p}\frac{|\bra{\bm{p},i}V\ket{B^{\alpha}}|^2}
    {(E_{p,i}+B^{\alpha})^2}\;  .
\end{align}
For an $s$-wave scattering, we have 
\begin{align}
    X^{\alpha}
    &= 
    \sum_{i}
    4\pi\sqrt{2\mu_{i}^{3}}
    \int_{E_{\text{th},i}}^{\infty} dE
    \sqrt{E-E_{\text{th},i}}
    \frac{|G^{\alpha}_{i}(E)|^{2}}
    {(E+B^{\alpha})^2}\;  ,
\end{align}
where we define $\bra{\bm{p},i}V\ket{B^{\alpha}}= G^{\alpha}_{i}(E_{p,i})$. This is the generalization of Eq.~\eqref{eq:Xexact}. The coupled channel scattering amplitude $t_{ij}(E)$ can be written as 
\begin{align}
    t_{ij}(E)
    &= 
    v_{ij}
    +
    \sum_{\alpha}
    \frac{G^{\alpha}_{i}(E)G^{\alpha}_{j}(E)}
    {E+B^{\alpha}} \nonumber \\
    &\quad +
    \sum_{k}
    4\pi\sqrt{2\mu_{k}^{3}}
    \int_{E_{\text{th},k}}^{\infty} dE^{\prime}
    \sqrt{E^{\prime}-E_{\text{th},k}}
    \frac{t_{ik}(E^{\prime})t_{kj}(E^{\prime})}
    {E-E^{\prime}+i\epsilon}\;  .
\end{align}
The summation in the second term is needed if there are other bound states than $\ket{B^{\alpha}}$. The compositeness $X^{\alpha}$ is then given by 
\begin{align}
    X^{\alpha}
    =& 
    \sum_{i}
    4\pi\sqrt{2\mu_{i}^{3}}
    \int_{E_{\text{th},i}}^{\infty} dE
    \frac{\sqrt{E-E_{\text{th},i}}}
    {E+B^{\alpha}}  
    \Biggl[
    t_{ii}(E)
    -\sum_{\beta\neq\alpha}
    \frac{|G^{\beta}_{i}(E)|^2}{E+B^{\beta}}
    \nonumber\\
    &-v_{ii} -\sum_{k}
    4\pi\sqrt{2\mu_{k}^{3}}
    \int_{E_{\text{th},k}}^{\infty} dE^{\prime}
    \sqrt{E^{\prime}-E_{\text{th},k}}
    \frac{|t_{ik}(E^{\prime})|^{2}}
    {E-E^{\prime}+i\epsilon}
    \Biggr]\; .
\end{align}
This is the generalization of Eq.~\eqref{eq:Xexactamp}. Again, the second term in the parenthesis vanishes if $\ket{B^{\alpha}}$ is the only bound state in the system. In the weak-binding limit $B^{\alpha}\ll E_{\rm typ}$, we have
\begin{align}
    X^{\alpha}
    \approx
    & 
    \sum_{i}
    4\pi\sqrt{2\mu_{i}^{3}}
    \frac{(g^{\alpha}_{0,i})^{2}}{\sqrt{E_{\text{th},i}+B^{\alpha}}}
    \; ,
\end{align}
where we define $g^{\alpha}_{0,i}\equiv G^{\alpha}_{i}(E=0)$.

It is instructive to concentrate on the lowest energy channel $\ket{\bm{p},1}$. Suppose that we integrate out the higher energy scattering channels by using the Feshbach projection method,\cite{Feshbach:1958nx,Feshbach:1962ut} and analyze the same system in the single-channel basis~\eqref{eq:completeness0} by identifying $\ket{\bm{p}}=\ket{\bm{p},1}$. Because the Feshbach projection does not alter the physical state, we have $\ket{B}=\ket{B_{\alpha}}$. This means that the compositeness $X$ is given by
\begin{align}
    X
    &= \int d\bm{p}|\bra{\bm{p}}\kket{B}|^{2}
    = \int d\bm{p}|\bra{\bm{p},1}\kket{B_{\alpha}}|^{2} 
    =X^{\alpha}_{1}\;  .
\end{align}
On the other hand, the elementariness $Z$ is given by
\begin{align}
    Z
    & 
    =1-X
    =Z^{\alpha}+\sum_{i=2}^{I}X_{i}^{\alpha}\;  .
    \label{eq:Zsingle}
\end{align}
Thus, we find that all the effect other than the scattering state $\ket{\bm{p},1}$ is included in the elementariness $Z$. This analysis explicitly demonstrates that the origin of $Z$ is not necessarily bare states, and the eliminated higher energy channels also contribute to $Z$. Even if we start from the coupled-channel model with no bare state $Z^{\alpha}=0$, the elimination of the coupled channels is translated as the elementary contribution in the single channel framework. In this sense, the elementariness $Z$ expresses something other than the model space of the scattering, as the CDD pole contribution.

\section{Compositeness of Unstable Resonances}\label{sec:resonance}

We now turn to the unstable resonances. First, we point out the fundamental difference between the resonances and the bound states. The probabilistic interpretation of the compositeness/elementariness of the bound states is guaranteed by 
\begin{itemize}
\item the normalization ($1=Z+X$) given in Eq.~\eqref{eq:normalization}, and 
\item the non-negativeness of the field renormalization constant in Eq.~\eqref{eq:Zdef}. 
\end{itemize}
The unity in the left-hand side of Eq.~\eqref{eq:normalization} follows from the normalization of the state vector of the physical bound state 
\begin{align}
    \bra{B}\kket{B}=1\;  .
\end{align}
This is possible because the bound state wave function is square integrable. In fact, denoting the eigenmomentum of the bound state as $p_{B}=i\kappa$ with a real and positive $\kappa$, we have the asymptotic behavior of the wave function at a large $|\bm{x}|$ as
\begin{align}
    \bra{\bm{x}}\kket{B}\to e^{ip_{B}|\bm{x}|}
    =e^{-\kappa |\bm{x}|} \; ,
    \label{eq:boundstate}
\end{align}
which vanishes exponentially in the limit $|\bm{x}|\to\infty$. The non-negativeness of $Z$ follows from the property $\bra{B}\kket{B_{0}}=\bra{B_{0}}\kket{B}^{*}$ as 
\begin{align}
    Z=\bra{B}\kket{B_{0}}\bra{B_{0}}\kket{B}=|\bra{B_{0}}\kket{B}|^{2}\;  .
\end{align}
In this way, the probabilistic interpretation of $Z$ and $X$ is a consequence of the normalizable state vector of the bound state.

For the resonances, the state vector is expressed by the Gamow vector $\ket{R}$. The Gamow vector is an improper vector and its norm is infinite. Intuitively, the generalization of Eq.~\eqref{eq:boundstate} may be obtained by the complex eigenmomentum $p_{R}=\alpha-i\beta$ with $\alpha>0, \beta>0$ as\footnote{The resonance pole should lie in the fourth quadrant in the complex $p$ plane which corresponds to the lower half of the second Riemann sheet of the complex energy plane.}
\begin{align}
    \bra{\bm{x}}\kket{R}\to e^{ip_{R}|\bm{x}|}
    =e^{\beta |\bm{x}|+i\alpha|\bm{x}|}\;  ,
\end{align}
which grows exponentially for $|\bm{x}|\to\infty$. This is formally understood as follows. If the state vector $\ket{R}$ is square integrable, the eigenvalue of the Hermite operators (such as the Hamiltonian) must be real, which contradicts to the complex energy of the resonance. Thus, the resonance state vectors cannot be square integrable by definition, and the normalization of the compositeness/elementariness is not guaranteed. This is an essential difference from the stable bound states.

In order to normalize the resonance state vector $\ket{R}$, we need to introduce the antiresonance state $\ket{\tilde{R}}=\ket{R^{*}}$ in the bi-orthogonal basis.\cite{PTP33.1116,Berggren:1968zz,Bohm:1981pv} The antiresonance is an eigenstate with a different boundary condition and is expressed by the conjugate pole at $\tilde{p}_{R}=-\alpha-i\beta$.\footnote{The antiresonance pole is located in the third quadrant in the complex $p$ plane which corresponds to the upper half of the second Riemann sheet of the complex energy plane.} Then the quantity $\bra{\tilde{R}}\kket{R}$ is bounded above, so we can normalize it as
\begin{align}
    \bra{\tilde{R}}\kket{R}=1\;  .
    \label{eq:normalizationres}
\end{align}
With the completeness relation, this leads to the generalization of the field renormalization constant $Z$ as
\begin{align}
    Z
    =\bra{\tilde{R}}\kket{B_{0}}\bra{B_{0}}\kket{R}\;  ,
    \label{eq:Zresonance}
\end{align}
which is in general complex, because $\bra{\tilde{R}}\kket{B_{0}}=\bra{B_{0}}\kket{R}\neq \bra{B_{0}}\kket{R}^{*}$. Thus, the interpretation of $Z$ is not straightforward. In the following we review recent attempts to extend the compositeness approach to the resonances. 

\subsection{Integration of the spectral density}

The generalization of the compositeness approach to the resonances is first proposed in Refs.~\citen{Baru:2003qq} and \citen{Hanhart:2011jz}. This method utilizes the spectral density\cite{Bogdanova:1991zz} which is defined as
\begin{align}
    w(E)
    =4\pi\sqrt{2\mu^3}
    \frac{\sqrt{E}
    |G(E)|^2}{(E+B)^2} \;  .
    \label{eq:spectraldensity}
\end{align}
For the bound state, the compositeness is given by the spectral density as 
\begin{align}
    1-Z
    =\int_{0}^{\infty}w(E)dE\;  .
    \label{eq:compositenessSD}
\end{align}
In this sense, the information of the structure of the bound state is included in the spectral density $w(E)$ defined on the real energy axis above the threshold. When there is an inelastic open channel at the energy below $E=-B$, the bound state becomes a quasi-bound state with a finite width. Reference~\citen{Baru:2003qq} suggested to include this effect through the Flatt\'e parametrization for the spectral density.\cite{Flatte:1976xu} Assuming the same functional form of Eq.~\eqref{eq:spectraldensity} for the resonances, the field renormalization constant $Z$ is obtained as a real number.\footnote{The denominator of Eq.~\eqref{eq:spectraldensity} is in fact $|E+B|^{2}$ for the bound state.} This method has been applied to study the structure of the scalar mesons $f_{0}(980)$ and $a_{0}(980)$\cite{Baru:2003qq} in the $\bar{K}K$ scattering and the $X(3872)$ resonance in the $D\bar{D}^{*}$ scattering.\cite{Hanhart:2011jz}

Strictly speaking, Eq.~\eqref{eq:spectraldensity} does not hold for the resonances, because of the normalization~\eqref{eq:normalizationres} which requires $|G(E)|^{2}\to [G(E)]^{2}$. On the other hand, the formulation is reduced to the bound state case when the decay width of the quasi-bound state is taken to be zero. Therefore, Eq.~\eqref{eq:spectraldensity} may be valid effectively for a narrow width state. This can justify the use of this strategy for the analysis of the narrow hadron resonances in Refs.~\citen{Baru:2003qq} and \citen{Hanhart:2011jz}.

\subsection{Field renormalization constant on the resonance pole}

In Subsec.~\ref{subsec:Zdef} we show that the compositeness is given by the binding energy and the transition form factor. These quantities are related to the position and its residue of the bound state pole in the scattering amplitude. Thus, we expect that the generalization to the resonances is accomplished by analyzing the properties of the resonance pole in the complex energy plane. To this end, we use the compositeness defined in the relativistic field theory with Yukawa coupling as\cite{Hyodo:2011qc}
\begin{align}
    X=1-Z
    &= \left.-g^{2}\frac{dG(W)}{dW}\right|_{W\to M_{B}} ,
    \label{eq:compositenessYukawa}
\end{align}
where $W$ is the total energy and $M_{B}$ is the mass of the bound state. The loop function $G(W)$ is given by
\begin{align}
    G(W)
    &= -\frac{1}{2\pi}
    \int_{s^{+}}^{\infty}ds^{\prime}
    \frac{\rho(s^{\prime})}{s^{\prime}-W^{2}-i0}
    +\text{(subtractions)}\; ,
\end{align}
where $\rho(s)$ stands for the phase space and $s^{+}$ is the squared threshold energy.\cite{Hyodo:2011ur} The coupling constant $g^{2}$ is determined by the residue of the pole of the amplitude on the real axis as
\begin{align}
    g^{2}
    &= 
    \lim_{W\to M_{B}}
    (W-M_{B})T(W)\; ,
\end{align}
where $T(W)$ is the scattering amplitude. Because the loop function $G(W)$ contains the energy denominator and the coupling constant $g^{2}$ expresses the transition from the bound state to the scattering state, Eq.~\eqref{eq:compositenessYukawa} is essentially same with Eq.~\eqref{eq:Xexact}.\footnote{The factorization of the coupling constant from the integration is the consequence of the scalar-type interaction in the Yukawa theory. With a different choice of the interaction Lagrangian, we would obtain a different expression of Eq.~\eqref{eq:compositenessYukawa}. This is the ``model dependence'' in the field theoretical formulation, which corresponds to the ambiguity of the choice of the potential $V$ in Eq.~\eqref{eq:Xgeneral}.} It is shown that $X\geq 0$ for the bound state,\cite{Hyodo:2011qc} so the probabilistic interpretation is ensured in this case.

A resonance state appears as a pole in the second Riemann sheet of the complex energy plane at $W=z_{R}\in \mathbb{C}$. The generalization of Eq.~\eqref{eq:compositenessYukawa} is given by
\begin{align}
    1-Z
    &=  \left.-g_{\rm II}^{2}\frac{dG_{\rm II}(W)}{dW}\right|_{W\to z_{R}} ,
    \label{eq:compositenessresonance}
\end{align}
where $G_{\rm II}(W)$ is the loop function in the second Riemann sheet and the coupling constant $g_{\rm II}^{2}$ is the residue of the amplitude in the complex plane,
\begin{align}
    g_{\rm II}^{2}
    &= 
    \lim_{W\to z_{R}}
    (W-z_{R})T(W)\;  .
\end{align}
The same result is obtained by the nonrelativistic separable potential with a sharp cutoff.\cite{Aceti:2012dd} The extension to the higher partial waves is also given in Ref.~\citen{Aceti:2012dd}.

The field renormalization constant $Z$ itself is well-defined even for the resonances, since it corresponds to the residue of the renormalized two-point function.\cite{Hyodo:2011qc} Moreover, the nonresonant scattering component does not contribute to the result because this approach utilizes the information on top of the resonance pole. However, the right-hand side of Eq.~\eqref{eq:compositenessresonance} is in general complex, because both $g_{\rm II}^{2}$ and $dG_{\rm II}(W)/dW|_{W\to z_{R}}$ are complex. This is a consequence of Eq.~\eqref{eq:Zresonance}. Thus, there is no fundamental problem of calculating Eq.~\eqref{eq:compositenessresonance}, while the interpretation of the obtained result is not straightforward. 

It is illustrative to present the formulation in the coupled-channel case, by comparing the notation in Subsec.~\ref{subsec:multi}. The normalization of the resonance vector~\eqref{eq:normalizationres} is expressed by the general relation in the coupled-channel amplitude\cite{Sekihara:2010uz,Hyodo:2011qc}
\begin{align}
    1&=\left.
    -\sum_{i,j}
    g_{i}g_{j}
    \left(
    \frac{dG_{i,{\rm II}}(W)}{dW}\delta_{ij}
    +G_{i,{\rm II}}(W)\frac{dV_{ij}(W)}{dW}G_{j,{\rm II}}(W)
    \right)\right|_{W\to z_{R}} ,
    \label{eq:WTidentity}
\end{align}
where $V_{ij}(W)$ is the interaction kernel. This is derived from the generalized Ward identity of the one-photon-attached scattering amplitude. Even though each component is complex, the total normalization is ensured by the gauge invariance. It is important to notice that Eq.~\eqref{eq:WTidentity} is evaluated at the resonance pole position $W=z_{R}$. When the interaction kernel is energy-independent, the second term vanishes and Eq.~\eqref{eq:WTidentity} reduces to the sum over the first term\cite{Aceti:2013jg}
\begin{align}
    1
    &=  -\left.\sum_{i}g_{i}^{2}\frac{dG_{i,{\rm II}}(W)}{dW}\right|_{W\to z_{R}}
    \quad \text{(energy-independent interaction)}
    \label{eq:sumrule}\; .
\end{align}
As discussed in Subsec.~\ref{subsec:interpretation}, the energy-dependence  of the interaction can be interpreted as the CDD pole contribution. Equation~\eqref{eq:sumrule} is therefore considered as the limit where all the CDD pole contribution is suppressed, so we define the compositeness as
\begin{align}
    X&=-\left.
    \sum_{i}
    g_{i}^{2}
    \frac{dG_{i,{\rm II}}(W)}{dW}
    \right|_{W\to z_{R}} 
    =
    \sum_{i}
    \int d\bm{p}\frac{\bra{\tilde{R}}V\ket{\bm{p},i}
    \bra{\bm{p},i}V\ket{R}}{(E_{p,i}-z_{R})^{2}} \; .
    \label{eq:compositenessresonancemulti}
\end{align}
All the contributions from the scattering state should be included in $X$. Thus, the elementariness of the resonance is identified as the rest contribution\cite{Sekihara:2012xp}
\begin{align}
    Z&=
    \sum_{n=1}^{N}
    \bra{\tilde{R}}\kket{B_{0,n}}\bra{B_{0,n}}\kket{R} 
    \nonumber \\
    &=
    1-
    \sum_{i=1}^{I}
    \int d\bm{p}
    \bra{\tilde{R}}\kket{\bm{p},i}\bra{\bm{p},i}\kket{R} 
    \nonumber \\
    &=
    -\left.
    \sum_{i,j}
    g_{i}
    G_{i,{\rm II}}(W)\frac{dV_{ij}(W)}{dW}G_{j,{\rm II}}(W)g_{j}
    \right|_{W\to z_{R}} ,
    \label{eq:elementarinessresonancemulti} 
\end{align}
where we have used the completeness of the bare states~\eqref{eq:completenessmulti}, definition of the compositeness~\eqref{eq:compositenessresonancemulti}, and the normalization~\eqref{eq:WTidentity}. We note that both $X$ and $Z$ are complex quantities, but its total normalization is ensured by Eq.~\eqref{eq:WTidentity}.

\subsection{Near-threshold resonances}

In the discussion of Subsec.~\ref{subsec:weakbinding}, we find that the structure of the bound states is model-independently determined in the weak-binding limit. We expect that a similar constraint may be derived in the analogous situation for the resonances, when the width and the excitation energy is small. The structure of such near-threshold resonances is studied in Ref.~\citen{Hyodo:2013iga}, using the effective range expansion.

In general, the behavior of the $s$-wave scattering amplitude $f(p)$ in the low momentum limit $p\to 0$ is determined by the effective range expansion given in Eq.~\eqref{eq:effectiverangeExp}. By truncating the expansion up to the $p^{2}$ order, the amplitude is completely specified by the scattering length $a$ and the effective range $r_{e}$. In this case, the poles of the amplitude are determined by $(a,r_{e})$ as\footnote{In Ref.~\citen{Hyodo:2013iga}, the scattering length is defined with opposite sign from Eq.~\eqref{eq:effectiverangeExp}. Here, we follow the convention given in Eq.~\eqref{eq:effectiverangeExp}, so the sign of $a$ in Eqs.~\eqref{eq:solution} and \eqref{eq:Zare} is opposite to that in Ref.~\citen{Hyodo:2013iga}. }
\begin{align}
    p^{\pm}
    = & \frac{i}{r_{e}}\pm \frac{1}{r_{e}}\sqrt{\frac{2r_{e}}{a}-1}
    \label{eq:solution}\;  .
\end{align}
For the single-channel scattering, the scattering length and the effective range are always real. The classification of the nature of these poles is given in Ref.~\citen{Ikeda:2011dx}. The near-threshold resonance can be realized with the negative effective range $r_{e}<0$. Thus, Eq.~\eqref{eq:solution} determines the scattering length and the effective range by the pole position of the near-threshold resonance. By eliminating $R$ from Eqs.~\eqref{eq:scatteringlength} and \eqref{eq:effectiverange}, we can express the field renormalization constant by $(a,r_{e})$ as
\begin{align}
    Z
    = & 1-\sqrt{
    1-\frac{1}{1-a/(2r_{e})}}
    \label{eq:Zare}\;  .
\end{align}
It is found that the compositeness $X=1-Z$ is purely imaginary, and normalized within $0<|X|<1$ for the resonances.\cite{Hyodo:2013iga} Because the normalization of the state vector is the crucial problem of the resonances, this may provide a hint for the interpretation of the compositeness of the resonances.

It is also worth noting that the single-channel near-threshold resonance is realized only with the negative effective range.\cite{Ikeda:2011dx,Hyodo:2013iga} As discussed in Subsec.~\ref{subsec:interpretation}, negativeness of the effective range is a measure of the contribution other than the scattering channel of interest. In other words, the existence of the near-threshold resonance itself implies the noncomposite nature of its structure.

\subsection{Applications to the hadron resonances}

In the recent works,\cite{Aceti:2012dd,Sekihara:2012xp,Xiao:2012vv,Aceti:2013jg,Hyodo:2013iga} the field renormalization constant $Z$ of the $s$-wave and $p$-wave hadron resonances has been evaluated by Eqs.~\eqref{eq:compositenessresonance}, \eqref{eq:elementarinessresonancemulti} and \eqref{eq:Zare}. We summarize the results of the field renormalization constant $Z$ in Table~\ref{tbl:Z}. We also show the absolute values $|Z|$ for reference. In some cases, the result depends on the cutoff of the loop function, reflecting the scheme-dependent nature of the field renormalization constant.

The field renormalization constant $Z$ measures the effect of the elementary contribution as the deviation from unity, while it is obtained as a complex number. A naive prescription for the interpretation is to take the absolute value.\cite{Aceti:2012dd,Sekihara:2012xp,Xiao:2012vv} Another prescription is to take the real part.\cite{Aceti:2013jg} In the examples shown in Table~\ref{tbl:Z}, two prescriptions provides roughly the same result, thanks to the relatively small imaginary part.

\begin{table}[bp]
\tbl{Field renormalization constant $Z$ of the hadron resonances evaluated on the resonance pole. The momentum cutoff $q_{\rm max}$ is chosen to be 1 GeV for the $\rho(770)$ and $K^{*}(892)$ mesons\cite{Aceti:2012dd,Xiao:2012vv}, 0.5 GeV for the $\Delta(1232)$ baryon, and 0.45 GeV for the $\Sigma(1385)$, $\Xi(1535)$, $\Omega$ baryons.\cite{Aceti:2013jg} }
{\begin{tabular}{@{}llllll@{}} \toprule
Baryons & $Z$ & $|Z|$ & Mesons & $Z$ & $|Z|$ \\ \colrule
$\Lambda(1405)$ higher pole (Ref.~\citen{Sekihara:2012xp}) & $0.00+0.09i$ & 0.09 & $f_{0}(500)$ or $\sigma$ (Ref.~\citen{Sekihara:2012xp})    & $1.17-0.34i$ & 1.22 \\
$\Lambda(1405)$ lower pole (Ref.~\citen{Sekihara:2012xp})  & $0.86-0.40i$ & 0.95 & $f_{0}(980)$ (Ref.~\citen{Sekihara:2012xp})                & $0.25+0.10i$ & 0.27 \\
$\Delta(1232)$ (Ref.~\citen{Aceti:2013jg})                 & $0.43+0.29i$ & 0.52 &
$a_{0}(980)$ (Ref.~\citen{Sekihara:2012xp})                & $0.68+0.18i$ & 0.70 \\
$\Sigma(1385)$ (Ref.~\citen{Aceti:2013jg})                 & $0.74+0.19i$ & 0.77 &
$\rho(770)$ (Ref.~\citen{Aceti:2012dd})                    & $0.87+0.21i$ & 0.89 \\
$\Xi(1535)$ (Ref.~\citen{Aceti:2013jg})                    & $0.89+0.99i$ & 1.33 &
$K^{*}(892)$ (Ref.~\citen{Xiao:2012vv})                    & $0.88+0.13i$ & 0.89 \\
$\Omega$ (Ref.~\citen{Aceti:2013jg})                       & $0.74$ & 0.74 &&&\\
$\Lambda_{c}(2595)$ (Ref.~\citen{Hyodo:2013iga})           & $1.00-0.61i$ & 1.17 &&& \\
 \botrule
\end{tabular} \label{tbl:Z}}
\end{table}%

We should again keep in mind that these numbers are not directly interpreted as the ``probability'' of the elementary component. This is clear because the result sometimes exceeds unity, as seen in the $\sigma$ meson and the $\Lambda_{c}(2595)$ baryon cases. On the other hand, it is clear that the magnitude of $Z$ (or $\re Z$) should reflect the amount of the elementary component,\cite{Aceti:2013jg} to some extent. It is an important future project to establish a firm interpretation of the field renormalization constant of the resonances.

\section{Other Approaches to the Hadron Structure}\label{sec:other}

We have been discussing the structure of hadrons from the viewpoint of the compositeness. This approach satisfies two conditions for a proper classification scheme summarized in Subsec.~\ref{subsec:conditions}; the compositeness is defined by the hadronic degrees of freedom and can be related to experimental observables. On the other hand, the extension to the resonances is not straightforward and we have not yet established a satisfactory method, as shown in Sec.~\ref{sec:resonance}. In the followings, we review the other approaches to study the structure of hadrons from different viewpoints. Since the different approaches shed light on the different aspects of the hadrons, the comparison of several approaches will be helpful to elucidate the nature of the exotic hadrons.

\subsection{Changing environment}\label{subsec:environment}

Let us consider a particle embedded in a surrounding environment. In general, when the environment (temperature, density, magnetic field, etc.) changes, the properties of the embedded particle (mass, width, etc.) will be modified, accordingly. The modification of the properties is driven by the interaction of the particle with the environment, so it reflects the structure of the particle. It is therefore a common exercise to investigate the response of the particle to the change of the environment, in order to pin down its structure. 

QCD is the gauge theory with the local color SU(3) symmetry, but it is known that the generalization to the SU($N_{c}$) theory with $N_{c}\to \infty$ offers new insight into the study of the strong interaction by the simple counting of the combinatorial factors.\cite{Hooft:1974jz,Witten:1979kh} One of the interesting consequences of the large $N_{c}$ limit is the simplified hadron structure in terms of the quarks and gluons. The $N_{c}$ counting tells us that a meson which survives in the large $N_{c}$ limit should be a pure $q\bar{q}$ state or a glueball.\footnote{See also a recent discussion on the tetraquarks in Ref.~\citen{Weinberg:2013cfa}. It is argued that a tetraquark state with decay rate suppressed by $1/N_{c}$ can survive in the large $N_{c}$ limit.} This leads to the $N_{c}$ scaling low of the mass and width of the $q\bar{q}$ meson as $m_{q\bar{q}}\sim \mathcal{O}(N_{c}^{0})$ and $\Gamma_{q\bar{q}}\sim \mathcal{O}(N_{c}^{-1})$. In the same way, the mass and width of the $N_{c}$-quark baryon $q^{N_{c}}$ behave as $m_{q^{N_{c}}}\sim \mathcal{O}(N_{c})$ and $\Gamma_{q^{N_{c}}}\sim \mathcal{O}(N_{c}^{0})$. 

The properties of a hadron resonance can be extrapolated to arbitrary $N_{c}$ by the theoretical models at $N_{c}=3$ with the general $N_{c}$ scaling of the model parameters. Comparing the model prediction with the $N_{c}$ scaling rules of the $q\bar{q}$ meson and the $N_{c}$-quark baryon, we extract the quark structure of hadrons. This strategy is initiated in Refs.~\citen{Pelaez:2003dy,Pelaez:2004xp,Pelaez:2006nj} for the $\sigma$ and $\rho$ mesons in the $\pi\pi$ scattering. The $N_{c}$ dependence is included in the low-energy constants of chiral perturbation theory, and the resonance pole position can be calculated for a given $N_{c}$. The deviation from the general scaling rule of the $q\bar{q}$ meson reflects the amount of the non-$q\bar{q}$ component of the resonance. The method is then applied to the axial vector mesons\cite{Geng:2008ag,Nagahiro:2011jn} and to the negative parity baryon resonances.\cite{Hyodo:2007np,Roca:2008kr} In addition to the $N_{c}$ extrapolation method using chiral models, it is also possible to extract the quark structure by estimating the magnitude of the $1/N_{c}$ corrections in the experimental observables.\cite{Nebreda:2011cp}

Another approach is to utilize the partial restoration of chiral symmetry. When chiral symmetry is partially restored, threshold enhancement is expected to occur in the $\pi\pi$ scattering amplitude in the scalar-isoscalar channel, as a consequence of the movement of the pole of the $\sigma$ meson. This is called the softening phenomena.\cite{Hatsuda:1999kd} It is shown that the behavior of the softening depends qualitatively on the structure of the $\sigma$ meson.~\cite{Hyodo:2010jp} Since chiral symmetry is expected to restore in the nuclear medium, this method opens a possibility of the experimental test for the structure of the $\sigma$ meson.

As a general remark on these approaches, we point out the nature transition during the extrapolation. The hadrons may not keep its original nature, after a long extrapolation from the physical world. For instance, a mesonic molecule state at $N_{c}=3$ can in principle be continuously extrapolated to the $q\bar{q}$-dominated state at very large $N_{c}$. In this case, the scaling analysis at very large $N_{c}$ does not make sense to determine the nature of the physical meson. This aspect is theoretically formulated as the mixing problem in the two-level model in Ref.~\citen{Nawa:2011pz}, where the condition for the nature transition is related to the geometrical structure in the complex parameter space. To avoid the ambiguity of the nature transition along with the extrapolation, the properties of the hadrons should be studied in the parameter region not very far from the real world.

\subsection{Spatial size}\label{subsec:size}

The spatial size is a basic quantity that characterizes the structure of a particle. First of all, there is no conceptual ambiguity in the definition of the size (the spatial extent of the wave function). Although it is not directly related to the internal structure of the hadrons, we can speculate the typical size from its construction. For instance, the hadronic molecule structure should have a larger size than the single-hadron state whose size is estimated by the energy scale of the color confinement ($1/\Lambda_{\rm QCD}\sim$ 1 fm). 

The electromagnetic form factors of the $\Lambda(1405)$ resonance are evaluated in Refs.~\citen{Sekihara:2008qk} and \citen{Sekihara:2010uz}. Unfortunately, the difficulty of the resonances also applies to the form factors. Because the form factor is defined as the matrix element of the electromagnetic current by the state vector, it is obtained as a complex number for the resonances. This causes the problem of the interpretation, but the magnitude of the mean squared radius indicates a larger spatial size of $\Lambda(1405)$ than the normal hadrons.\cite{Sekihara:2008qk,Sekihara:2010uz}

Recently, a novel method to obtain a real-valued size of the resonances is proposed through the finite volume effect.\cite{Sekihara:2012xp} A general discussion on the properties of a stable bound state in a finite box is given in Ref.~\citen{Luscher:1985dn}. The mass shift of the bound state due to the finite volume effect is related to the coupling constant of the bound state to the scattering state in the infinite volume. This method is extended to the shift of the resonance pole position when a closed channel is put into a finite box, and the size is estimated as a real number.\cite{Sekihara:2012xp} We note that the finite volume effect provides a new definition of the ``size'' of the resonances, which does not exactly coincide with that defined from the complex form factor. In addition, the applicability of the method is limited to the spatial size of the closed channels. It is however important to extract the real-valued information of the size of the resonances. The result indicates a larger size of $\Lambda(1405)$, in accordance with the form factor approach.

\subsection{Production mechanism}\label{subsec:exp}

As emphasized in Subsec.~\ref{subsec:conditions}, the structure of the hadrons should be eventually examined by the experimental observables. One way is to focus on the production mechanism, since different structures may be produced in different ways. A study in this direction is performed by focusing on the hadronization process in the heavy ion collisions.\cite{Cho:2010db,Cho:2011ew} It is shown that the production yield of a hadron in the heavy ion collisions depends qualitatively on the structure of the hadron (multiquark or hadronic molecule). This study therefore provides an experimental verification for the structure of hadrons. It should be noted that the definition of the hadron structure in this approach is based on the hadronization mechanism, which does not exactly correspond to the definitions in other approaches. Again, comparison with different approaches will help elucidating the structure of the exotic hadrons.

\section{Summary}	

The purpose of this paper is to clarify the subtle issues in the study of the structure of the hadron resonances, and to present the possible solution to this problem. Based on a comprehensive discussion on the difficulty of the definition of the hadron structure, we conclude that the desired framework for the hadron structure should be given with the hadronic degrees of freedom and possibly in a model-independent manner. The applicability to the resonances with a suitable interpretation is also necessary. It is shown that the compositeness approach with the field renormalization constant is a good candidate for this strategy; the compositeness is well defined for the stable bound states, and is model-independently related to the experimental observables in the weak-binding limit. We would like to emphasize the following (not well-known) aspects of this approach.
\begin{itemize}

\item The normalization of the state vector $\bra{B}\kket{B}=1$ and the non-negativeness of the field renormalization constant $Z\geq 0$ are crucial for the probabilistic interpretation of the compositeness/elementariness. 

\item The compositeness of the bound state with an arbitrary binding energy depends on the choice of the basis of the bare Hamiltonian (the way to decompose the full Hamiltonian $H$ into $H_{0}$ and $V$). In other words, the compositeness is a scheme-dependent quantity. This scheme dependence vanishes in the weak-binding limit, where the compositeness is model-independently related to the experimental observables.

\item The elementariness is interpreted as the fraction of the CDD pole contribution which is anything other than the scattering states in the given model space. Once the model space is specified, the origin of the elementariness cannot be known. Because the elementariness is expressed by the energy dependence of the potential, it can be traced back to the consequence of the elimination of the coupled-channel effect in a larger model space. In the weak-binding limit, the magnitude of the negative effective range reflects the elementariness. 

\end{itemize}

We discuss the generalization of the compositeness approach to the resonances. It is shown that the normalization of the resonance vector can be expressed by the generalized Ward identity. On the other hand, the field renormalization constant becomes a complex number, reflecting the bi-orthogonal nature of the resonance state vectors. It is still an open problem to establish a firm framework which is capable of treating the resonances with a suitable interpretation of the result. We hope that the discussion in this paper serves as a useful cornerstone for future developments in the study of the structure of the hadron resonances.

\section*{Acknowledgments}

The author is grateful to all the collaborators of the studies for the hadron structure. Fruitful discussions with many colleagues in various occasions are deeply appreciated. Among others, the author would like to sincerely express his gratitude to two mentors on this subject, Daisuke Jido and Atsushi Hosaka. The author thanks the Yukawa Institute for Theoretical Physics at Kyoto University. Discussions during the YITP workshop YITP-W-12-19 on ``Resonances and non-Hermitian systems in quantum mechanics'' were useful to complete this work. This work is partly supported by the Grant-in-Aid for Scientific Research from MEXT and JSPS (Grants No. 24105702 and No. 24740152).



\end{document}